\documentclass{aa}  

\usepackage{graphicx}
\usepackage{chemist}
\usepackage{commandes_GMM}
\usepackage[version=3]{mhchem}

\usepackage{txfonts}
\usepackage{eufrak}
\usepackage{euscript}
\usepackage{array,multirow,makecell}
\usepackage{tikz}
\begin{document} 
\newcommand {\be}{\begin{equation}} 
\newcommand{\ee}{\end{equation}}

   \title{Consistent transport properties in multicomponent  two-temperature magnetized plasmas:}

   \subtitle{Application to the Sun chromosphere}

   \author{Q. Wargnier
          \inst{1}
          \and
          A. Alvarez Laguna
          \inst{1,2}
          \and
          J. B. Scoggins
          \inst{3}    
          \and
          N. N. Mansour
          \inst{4}
          \and
          M. Massot
          \inst{1}
          \and
          T. E. Magin   
           \inst{3}  
          }

   \institute{Centre de Math\'ematiques Appliqu\'ees, Ecole Polytechnique, CNRS, Palaiseau, France
	    \and 
             Laboratoire de Physique des Plasmas, Ecole polytechnique, CNRS, Palaiseau, France
              \and
             Aeronautics and Aerospace Department, von Karman Institute for Fluid Dynamics,  Rhode-Saint-Gen\`ese, Belgium
             \and
             NASA Ames Research Center,  Moffett Field, California, USA
                    }

   \date{Received ??; accepted ??}
 
 
  \abstract
   {}
   {}
  {}
   {}
   {}

 \abstract{}{A fluid model is developed for multicomponent two-temperature magnetized plasmas in chemical non-equilibrium from  the  partially-  to  fully-ionized  collisional  regimes.  We focus on transport phenomena aiming  at  representing the chromosphere of the Sun. }{Graille $et~al.$ [M3AS 19(04):527-599, 2009]  have derived an asymptotic fluid model for multicomponent plamas from kinetic theory, yielding a rigorous description of the dissipative effects. The governing equations and consistent transport properties are obtained using a multiscale Chapman-Enskog perturbative solution to the Boltzmann equation  based on a non-dimensional analysis. The mass disparity between the electrons and heavy particles  is accounted for, as well as the influence of the electromagnetic field.  We couple this model to the Maxwell equations for the electromagnetic field and derive the  generalized Ohm's law for multicomponent plasmas. The model inherits a well-identified mathematical structure leading to an extended range of validity for the Sun chromosphere conditions. We compute  consistent transport properties by means of a spectral Galerkin method using the Laguerre-Sonine polynomial approximation. Two non-vanishing polynomial terms are used when deriving the transport systems for electrons, whereas only one  term is retained for heavy particles.  In a simplified framework where the plasma is  fully ionized, we compare the transport properties for the  Sun chromosphere to conventional expressions for magnetized plasmas due to Braginskii, showing a good agreement between both results.}
 {For more  general partially ionized conditions, representative of the Sun chromosphere, we compute the muticomponent transport properties corresponding to the species diffusion velocities, heavy-particle and electron heat fluxes, and viscous stress tensor of the model,  for a  Helium-Hydrogen mixture in local thermodynamic  equilibrium. The  model is assessed for the  3D radiative magnetohydrodynamic simulation of a pore, in the highly turbulent upper layer of the solar convective zone. The  resistive term is found to dominate mainly the dynamics of the electric field at the pore location. The battery term for heavy particles appears to be higher at the pore location and at some intergranulation boundaries.}{}
   \maketitle
%

\section{Introduction}
\label{sec:introduction}

The lower atmosphere of the Sun is a complex and dynamic layer where the plasma is found in a wide range of different regimes -- from weakly-ionized and non-magnetized at the bottom of the photosphere to fully ionized and magnetized at the top of the transition region. In the Sun chromosphere, the pressure varies from a thousand pascals just above the photosphere to a few pascals in the high chromosphere \cite{Vernazza}. Similarly, the magnitude of the magnetic field is large in active regions, around thousands of gauss in sunspots, whereas it is just a few gauss in quiet-Sun regions \cite{wiegelmann}. It is still nowadays a great challenge to develop a unified model that can be used for both  partially- and fully-ionized regimes under the large disparity of plasma parameters in the chromosphere.

The study of partially-ionized plasmas in the presence of a magnetic field, such as in prominences and  the lower atmosphere of the Sun, demands models  that are beyond the ideal single-fluid MagnetoHydroDynamic (MHD) description. Phenomena necessary to fully understand the behavior of plasmas in the Sun chromosphere, such as Cowling's resistivity, thermal conduction, heating due to ion-neutral friction, heat transfer due to collisions, charge exchange collisions and ionization energy losses, are usually disregarded in ideal MHD models, or only described by means of ad-hoc terms.

Two types of fluid models to study the Sun chromosphere are found in the literature. First, the single-fluid MHD description considers the plasma as a conducting fluid in the presence of a magnetic field. It has the main drawback of assuming thermal equilibrium conditions, where all the species  are considered to be at the same temperature. This model is assumed to be valid in the lower Sun atmosphere, allowing us to study the formation of magnetic field concentrations at the solar surface in sunspots, magnetic pores, and the large-scale flow patterns associated with them \cite{Kosovichev}. It is also used for simulating the lower part of the atmosphere of the Sun, $e.g.$, incorporating subgrid-scale turbulence models for the transport of heat and  electrical resistivity \cite{Irina}.  The full MHD equations  are solved in \cite{sykora} accounting for non-grey radiative transfer and thermal conduction outside local thermodynamic equilibrium, in order to study the effects of the partial ionization of the Sun chromosphere.

Second,  multi-fluid MHD models  have been used  more recently to represent the non-equilibrium conditions of the chromosphere, based on continuity, momentum, and energy conservation equations for each species considered in the mixture \cite{khomenko2,Khomenko, laguna, laguna2018, Leake12, Ni2018}. However, these models lead to very stiff systems that are difficult to solve numerically as they exhibit characteristic times that range from the convective and diffusive times of each fluid down to the collisional and chemical kinetics time scales \cite{laguna}. Leake et al. \cite{Leake12} performed a multi-fluid simulation of magnetic reconnection for a weakly ionized reacting plasma, with a particular focus on the solar chromosphere, by considering collisional transport,  chemical reactions between  species, as well as radiative losses.   
Braginskii~\cite{Braginskii} has derived rigorous expressions for the transport properties of fully-ionized plasmas starting from the Boltzmann equation. Khomenko et al. \cite{Khomenko} proposed a model for the description of a multi-component partially ionized solar plasma. Deriving rigorous transport properties for such multi-fluid model is complex, and so far, the theory has not yet been developed to the same level of accuracy as Braginskii's.

In this paper, we propose a novel approach for studying the Sun chromosphere that is neither a single-fluid  nor  a multi-fluid MHD model, but a multi-component drift-diffusion model derived  by Graille et al. from kinetic theory~\cite{graille}. This approach is able to capture most of the multi-fluid phenomena, i.e., different velocities between species, collisional exchange of mass momentum and energy, chemical reactions, thermal non-equilibrium, etc. Besides, the system of equations is less stiff as it solves for only one momentum equation, like in the single-fluid MHD approach. The transport properties are retrieved through a generalized Chapman-Enskog solution to the Boltzmann equation that uses a multiscale perturbation method. These developments lead to a model with an extended range of validity from partially- to fully-ionized plasmas, with or without the presence of magnetic field. We couple this model to the Maxwell equations for the electromagnetic field  and derive the corresponding generalized Ohm's law for multicomponent plasmas. As in Braginskii's theory, our model includes  anisotropy in the transport properties of electrons, that is created by the magnetic field.  These properties are computed by solving for the integro-differential systems presented by Scoggins et al.~\cite{scoggins}. We use a spectral Galerkin method based on the Laguerre-Sonine polynomial approximation previously  studied in depth for various applications~\cite{bruno,ferzigerkapper,kinetictheory,Transportproperties,zhdanov,magin04,devoto,capitelli}. The transport systems are  implemented in the Mutation++ library that compiles state-of-the-art transport collision integral data for the different pairs of species in the mixture~\cite{mutation}. In a simplified framework where the plasma is  fully ionized, we compare the transport properties for the  Sun chromosphere to the conventional expressions for magnetized plasmas due to Braginskii. For more  general partially ionized conditions representative of the Sun chromosphere, we compute the muticomponent transport properties corresponding to the species diffusion velocities, heavy-particle and electron heat fluxes, and viscous stress tensor,  for a  Helium-Hydrogen mixture in local thermodynamic  equilibrium. Finally, the  model is assessed for the  3D radiative magnetohydrodynamic simulation of a pore, in the highly turbulent upper layer of the solar convective zone.  We compute the thermal conductivity, electrical conductivity,  species diffusion coefficients, and the components of the generalized Ohm's law and conclude on the importance of the contribution of its components, in particular, of the resistive and battery terms.

The structure of the paper is as follows. In section \ref{sec:sec2}, the non-dimensonal analysis used for the generalized Chapman-Enskog expansion is presented, together with the multi-component drift-diffusion model for    two-temperature magnetized plasmas, the transport fluxes, and the generalized Ohm law. In section \ref{methodology}, one describes the mixture considered, the conditions representative of the Sun chromosphere and the method used for computing the transport properties. In section \ref{sec:sec5}, we verify the  model
proposed on a fully-ionized case by comparing the results with  those obtained by means of Braginskii's theory. Finally, in section \ref{sec:sec3}, we discuss all the transport properties for a partially ionized case. Additionally, we compute the transport properties and the components of the generalized Ohm's law for  3D radiative  MHD simulations of a pore in the low Sun atmosphere.

\section{Drift-diffusion model for multicomponent plasmas}
\label{sec:sec2}
In this section, we present the  multi-component drift-diffusion model for    two-temperature magnetized plasmas. It was derived from  kinetic theory by Graille et al.~\cite{graille} as a generalized Chapman-Enskog solution to the Boltzmann equation, using a multi-scale perturbation method based on a non-dimensional analysis.  Additionally, we compare this model to the multi-fluid description widely used for the Sun chromosphere. 

\subsection{Multi-scale analysis of the Boltzmann equation}
We consider a multicomponent plasma composed of electrons, denoted here by the index $\elec$, and heavy-particles (atoms and molecules, neutral or ionized), denoted the subscript $\heavy$. The species are assumed to be point particles, neglecting their internal energy. We combine the equations derived in \cite{graille,chimie} for the fully magnetized case and the Maxwellian regime for reactive collisions. 

In order to apply the Chapman-Enskog method, Graille et al.~\cite{graille} perform a multi-scale analysis on the non-dimensional Boltzmann equations for electrons and heavy species. The order of magnitude of the different terms in the Boltzmann equation is studied by choosing carefully reference quantities. In the asymptotic fluid limit, the Knudsen number is assumed to be of the same order of magnitude of the square root of the mass ratio between electrons and heavy particles, defined as \smash{$\epsilon=\sqrt{{\me}/{\mh}}$}, where $\me$ and $\mh$ are the mass of electrons and heavy particles respectively. This small parameter drives thermal nonequilibrium between the electron and heavy-particle baths. In the strongly magnetized regime, the Hall parameter is assumed to scale as $\epsilon^0=1$. The species distribution functions are expanded in the multiscale perturbation parameter  $\epsilon$ following Enskog's approach. As opposed to Braginskii, no  assumption is made a priori on the zero-order distribution function. The asymptotic analysis of the Boltzmann equation is performed at successive orders of $\epsilon$. The main results  occurring at different time-scales  are summarized in Table \ref{tab:CE}. 
 \begin{table}[ht]
 	\centering\footnotesize
 	\caption{Time scales hierarchy and macroscopic equations derived using the Chapman-Enskog method~\cite{graille}}
 	\label{tab:CE}
 	\renewcommand*{\arraystretch}{1.5}
 	\begin{tabular}{@{}p{0.5cm}p{1.cm}p{3cm}p{3cm}@{}}
 		\bfseries{Order} & \bfseries{Time} & \bfseries{Heavy particles} &\bfseries{Electrons} \\ \hline
 		$\epsilon^{-2}$ & $\refte$ & &Thermalization at $\tempe$ \\ 
 		$\epsilon^{-1}$ & $\refth$ &Thermalization at $\temph$ &\\
 		$\epsilon^{0}$ & $\reft$ &Euler & $0^{th}$-order drift-diffusion\\
 		$\epsilon$ &$\reft/\epsilon$ &Navier-Stokes &$1^{st}$-order drift-diffusion
 	\end{tabular}
 \end{table}
 
In time scales of order $\refte$, the electron population thermalize at the temperature $\tempe$. The electron distribution function is a Maxwell-Boltzmann distribution obtained by solving the electron Boltzmann equation at the order $\epsilon^{-2}$. At order $\epsilon^{-1}$ that corresponds to the time scale $\refth$, heavy particles thermalize at  temperature $\temph$. At the zeroth order \smash{$\epsilon^{0}$} that corresponds to the convective time scales, Euler equations for heavy particles and zero-order drift-diffusion equations for electrons are obtained. Finally, at order $\epsilon$, corresponding to the diffusive time scale, we obtain Navier-Stokes equations for heavy particles and first-order drift-diffusion equations for electrons.  It is important to mention that in Braginskii's approach, the macroscopic equations are retrieved by taking moments of the Boltzmann equation by assuming directly that the zero-order Maxwell-Boltzmann distributions. Only a correct scaling deduced from dimensional analysis can yield a sound multicomponent  treatment.


\subsection{Multi-component equations}
\label{sec:governing}
We denote by symbol $\lourd$ the set of indices for the heavy particles of the mixture considered. First, the mass conservation equations for electron and heavy particles are, as follows,
\begin{equation}
\dt\rhoe + \dx\dscal\Bigl[\rhoe(\vitesse+\Ve) \Bigr] =  \omegae ,
\label{eq:rhoe}
\end{equation}
\begin{equation}
\dt\rhoi + \dx\dscal\Bigl[\rhoi(\vitesse + \Vi) \Bigr] = \omegai,\quad i \in \lourd.
\label{eq:rhoh}
\end{equation}
Here, $\rhoe$ is the density of electron,  $\rhoi$ is the density of heavy particle $i \in \lourd$,  $\vitesse$ is the heavy particle hydrodynamic velocity that has been chosen as the velocity reference frame, $\Ve$ is the electron diffusion velocity, $\Vi , i \in \lourd$ is the heavy diffusion velocity of heavy particle $i$ in the heavy hydrodynamic reference frame, such that $\sum_{i\in\lourd}\rhoi\Vi=0$. Quantities $\omegae$ and $\omegai$, with $i \in \lourd$, are respectively the chemical production rates of electrons and heavy particles. 

Second, the one momentum equation Eq.~\eqref{eq:momentum} for all the particles within plasma is
\begin{equation}
		\dt(\rhoH\vitesse)+\dx\dscal\left[\rhoH\vitesse\ptens\vitesse+ \pression\identite\right]=-\dx\dscal\visqueux[\heavy]+ \ntot\qtot\E 
		+\courantel\pvect\B.\label{eq:momentum}
\end{equation}		
Here, $\pression= \pree+ \prh$ is the total pressure which is the sum of the partial pressure of electron $\pree$ and heavy particle $\prh$,  $\visqueux[\heavy]$ is the viscous stress tensor.  $\rhoH$ is the total density of heavy particles, where $\ntot=\nee+\nH$ is the density ($\nee$ the number of electron and $\nH$ the total number of heavy particle per unit of volume). $\ntot \qtot$ is the total charge of the system defined by $\ntot\qtot=\nee \qe + \sum_{i \in \lourd}\ni\qi$, $\E$ is the electric field, and $\courantel$ is the total current density defined as
\be
\courantel=\ntot\qtot \vitesse+ \JJe +\JJi[\heavy]  = \ntot \qtot \vitesse+ \nee \qe \Ve + \sum_{i \in \lourd}\ni\qi\Vi,
\label{eq:totalcurrent}
\ee
where $\JJi[\heavy]$ is the heavy-particle conduction current density, $\JJe$ is the electron conduction current density, $\courantel$ is the total current density,  and $\B$ is the magnetic field.

Third, the two equations (\ref{eq:rhoeee}) and (\ref{eq:rhoheh}) for the thermal energies of electrons and heavy particles are	
\begin{multline}
\dt(\rhoe\energiee)+\dx\dscal\left[\rhoe\energiee\vitesse\right]+\pree\dx\dscal\vitesse+\\
\dx\dscal\heate=\JJe\dscal\Ep-\deltaEho-\deltaEhu+ \Omega_{\elec},
\label{eq:rhoeee}
\end{multline}
\begin{multline}
\dt(\rhoH\energiei[\heavy])+\dx\dscal\left[\rhoH\energiei[\heavy] \vitesse\right] +(\pri[\heavy]\identite+\visqueux[\heavy])\pmat\dx \vitesse+ \\
\dx\dscal\heati[\heavy] =\JJi[\heavy]\dscal\Ep+\deltaEho + \deltaEhu+\Omega_{\heavy},
\label{eq:rhoheh}
\end{multline}
where $\rhoe\energiee$ and $\rhoH\energiei[\heavy]$ are the internal energies of electron and heavy-particles respectively, and $\heate$ and $\heati[\heavy]$ are the electron and heavy particle heat flux, respectively. $\Ep=\E+\vitesse\pvect\B$ is the electric field in the heavy particle reference france, $\deltaEho$ and $\deltaEhu$ are the relaxation terms at order $\epsilon^0$ and $\epsilon$, respectively.   $\JJi[\heavy]\dscal\Ep$ and $\JJe\dscal\Ep$ are respectively the power that is developed by the heavy particle and electron current density, and $\Omega_{\elec}$ and $\Omega_{\heavy}$ are respectively the energy production rate for electrons and heavy particles. 

By summing the equations of internal energies, i.e., Eq.~\eqref{eq:rhoeee} and Eq.~\ref{eq:rhoheh}, and the equation of kinetic energy, the equation of total energy can be obtained, as follows,
\be	
\dt\energie+\dx\dscal\left[\left(\energie+\pression\right)\vitesse\right]+\dx\dscal(\visqueux[\heavy]\dscal\vitesse)+\dx\dscal\left(\heate+\heati[\heavy]\right)=\courantel\dscal\E,
\label{eq:totalEnergy}
\ee
where $\energie=\rhoe\energiee+\rhoH\energiei[\heavy]+1/2\rhoH \vitesse^2$, $\courantel\dscal\E$ is the power developed by the electromagnetic field.

The system of equations (\ref{eq:rhoe},\ref{eq:rhoh},\ref{eq:momentum},\ref{eq:rhoeee},\ref{eq:rhoheh},\ref{eq:totalEnergy}) is coupled to the set of Maxwell's equations (\ref{eq:maxwell}) :
\begin{equation}
\begin{aligned}
&\dx\dscal\E = \frac{\ntot\qtot}{\epsilon_0},   \\
&\dx\dscal\B=0 , \\
&\dt\B=-\dx\pvect \E, \\
&\dx\pvect \B=\mu_0 \courantel + \mu_0\epsilon_0\dt \E 
\end{aligned}	
\label{eq:maxwell}
\end{equation}
where $\epsilon_0$ is the vacuum permittivity and $\mu_0$ the vacuum permeability. 

The electron transport fluxes such as the electron diffusion velocity $\Ve$, the electron heat flux $\heate$, the electron current density $\JJe$ are composed of two terms: 1- a term at the convective time scale, at order $\epsilon^0$, corresponding to the Euler equations for heavy species, and 2- a term which is a first order correction, at order $\epsilon$, at the dissipative time scale corresponding to the Navier-Stokes equations for heavy species. 

Similarly, the heavy transport fluxes such as the heavy-particle diffusion velocity $\Vi,i\in\lourd$, the heavy-particle heat flux $\heati[\heavy]$, the viscous stress tensor  $\visqueux[\heavy]$, the heavy particle current density $\JJi[\heavy]$ are defined at the dissipative time scale, at the order $\epsilon$ of the generalized Chapman-Enskog expansion. In \cite{graille}, the momentum equation for electrons is vanishing in the generalized Chapman-Enskog expansion, due to the nondimensional analysis. Electrons do participate in the momentum balance through the total pressure gradient and Lorentz force. Besides, electrons do not participate in the stress tensor due to the small mass disparity between electrons and heavy particles.

The governing equations  (\ref{eq:rhoe},\ref{eq:rhoh},\ref{eq:momentum},\ref{eq:rhoeee},\ref{eq:rhoheh},\ref{eq:totalEnergy},\ref{eq:maxwell}) differ from the multi-fluid models used for partially ionized plasmas. Whereas multi-fluid models consider one hydrodynamic velocity for each species, here we consider one hydrodynamic velocity for the heavy species while each species diffuse in this reference frame.

In addition, the structure of the governing equations is symmetrizable hyperbolic, which can be regarded as an important property for the numerical discretization of the system. Nevertheless, it is necessary to close the model by computing the transport properties. This computation is presented in section \ref{sec:sec3} for a Helium-Hydrogen mixture. Additionally, by using the definition of the total current density $\courantel$ and the Maxwell equations (\ref{eq:maxwell}), a generalized Ohm's law for this particular model is derived in Section \ref{sec:sec4}.

\subsection{Transport fluxes for heavy particles}
With the same formalism that is used in Graille et al.~\cite{graille}, we introduce some extra notations in order to express the anisotropic transport properties in the presence of a magnetic field. First,  a unit vector for the magnetic field $\Bnorme=\B/|\B|$ is defined and also three direction matrices
\begin{equation*}
\Mpa = \Bnorme\ptens\Bnorme,
\quad
\Mpe = \identite - \Bnorme\ptens\Bnorme,
\quad
\\ \Mt = \begin{pmatrix}
0 & -\normBnorme[3] & \normBnorme[2] \\
\normBnorme[3] & 0 & -\normBnorme[1] \\
-\normBnorme[2] & \normBnorme[1] & 0
\end{pmatrix}
\end{equation*}
so that we have for any vector $\X$ in three dimensions
\begin{equation*}
\Xpa=\Mpa\X = \X\dscal\Bnorme\;\Bnorme,
\qquad
\Xpe=\Mpe\X = \X - \X\dscal\Bnorme\;\Bnorme,
\end{equation*}
\begin{equation*}
\Xt=\Mt\X = \Bnorme\pvect\X.
\end{equation*}
In the $(\X,\Bnorme)$ plane, the vector $\Xpa$ is the component of $\X$ that is parallel to the magnetic field and $\Xpe$ is the perpendicular component. Therefore, we have $\X=\Xpa+\Xpe$. The vector $\Xt$ lies in the direction transverse to the $(\X,\Bnorme)$ plane.
The three vectors $\Xpa$, $\Xpe$, and $\Xt$ are then mutually orthogonal. The anisotropic transport coefficients are expressed by means of the matrix notation
\be
\bar{\bar{\mu}}=\mu^\pa\Mpa+\mu^\pe\Mpe+\mu^\tran\Mt
\ee
If the transport coefficients are identical in the parallel and perpendicular directions, $\mu^\pa=\mu^\pe$, and vanish  in the transverse direction, $\mu^\tran=0$ an isotropic system is obtained. \\
In the governing equations presented in section (\ref{sec:governing}), the heavy particle transport fluxes are
\begin{center}
	$
\visqueux[\heavy],\quad \Vi, i\in\lourd ,\quad \heati[\heavy].
$
\end{center}
First, the viscous stress tensor $\visqueux[\heavy]$ is defined as 
\be
\visqueux[\heavy]=-\etah\left(\left[\dx\vitesse+(\dx\vitesse)^{\top}\right]-\frac{2}{3}\left(\dx\dscal\vitesse\right)\identite\right),
\label{eq:viscous}
\ee
where $\etah$ is the viscosity of heavy particles.
Then, the heavy particle diffusion velocity $\Vi, i \in \lourd$ is defined as 
\be
\Vi= - \sum_{j\in\lourd}\Dij\left(\djj +\chihj\glogTh\right),\quad i \in \lourd,
\label{eq:Vi}
\ee
where $\Dij$ is the multicomponent diffusion coefficient of heavy particles, $\djj$ is the diffusion driving force of the particle $j \in \lourd $ that is interacting with the heavy particle $i \in \lourd$, and $\chihj, j\in\lourd$ is the heavy thermal diffusion ratio. The diffusion driving force $\djj$ is defined as
\be
\djj=\frac{1}{\prh}\left(\dx \prj-\nj \qj\Ep- \nj\Fje \right),\quad j \in \lourd.
\label{eq:heavydiffusiondrivingforce}
\ee
The latter is composed of three forces: 1- the force due to the gradient of the partial pressure $\dx \prj, j \in \lourd$, 2- the Lorentz force and 3- $\Fje, j\in\lourd$ that is an average electron forces acting on the heavy particle $j$. $\Fje$ belongs to the category of diffusion driving forces and allows for a coupling between the heavy particles and electrons \cite{graille}. This average force is defined as
\be
\Fje=-\frac{\pree}{\nj}\malphaej\de-\frac{\pree}{\nj}\mchiej\glogTe,\quad j \in \lourd.
\ee
where $\malphaej, j \in \lourd$ and $\mchiej, j \in \lourd$ are anisotropic transport coefficients. Finally, the heavy particle heat flux reads 
\be
\heati[\heavy]= -\lambdaH\dx \temph+\prh\sum_{j\in\heavy}\chihj\Vi[j]+\sum_{j\in\lourd}\rhoi[j] h_{j}\Vi[j]
\label{eq:heath}
\ee
where $\lambdaH$ is the heavy thermal conductivity and $\rhoi[j]h_{j}$ is the enthalpy of heavy particle $j\in\lourd$.  The second term of Eq. \eqref{eq:heath} can be seen as a friction term between the species $i$ and $j$. \\

In the previous transport fluxes, some of the usual terms can be identified. The viscous stress tensor for heavy particles Eq. (\ref{eq:viscous}) is proportional to the strain tensor. Similarly, the first term of $\Vi$ in Eq. \eqref{eq:Vi} is a generalized Fick's law where the flux is proportional to the diffusion driving force $\djj, j \in \lourd$. Also, the first term of the heavy particle heat flux $\heati[\heavy]$ in Eq. \eqref{eq:heath} is the usual Fourier's law. Besides, $\Vi$ includes a term that is proportional to $\glogTh$. This term is known as the Soret effect, highly described in \cite{Transportproperties}.

In summary, the transport coefficients for heavy particles to be computed in the following sectionss are
\begin{center}
	$
	\etah,\quad \Dij,\quad \chihj,\quad \lambdaH,    \quad i,j\in\lourd.
	$
\end{center}
In addition, the anisotropic transport coefficients associated to the coupling terms between electron and heavy particles are
\begin{center}
$\malphaej, \quad \mchiej, \quad j \in \lourd$.
\end{center}

\subsection{Transport fluxes for electrons}
The electron transport fluxes are
\begin{center}
	$
	\Ve \quad\text{and}\quad \heate.
	$
\end{center}
The electron diffusion velocity is defined as
\be
\Ve= -\mDee\left(\de+\mchie\glogTe\right)+\sum_{i\in\lourd}\malphaei\Vi,
\label{eq:electrondiffusionvelocity}
\ee
where $\mDee$ is the tensor of the diffusion coefficient of electrons, $\de$ is the electron diffusion driving force and $\mchie$ is the electron thermal diffusion ratio. The electron diffusion velocity $\Ve$  is splitted into two terms : 1- the terms proportional to $\de$ and to $\glogTe$ are at order $\epsilon^0$, at the heavy particle convective timescale and 2- the terms proportional to $\Vi,i \in\lourd$ are at order $\epsilon$ at the heavy particle dissipative timescale. The electron diffusion driving force is defined as 
\be
\de=\frac{1}{\pree}\left(\dx \pree-\nee \qe\Ep \right).
\label{eq:electrondiffusiondrivingforce}
\ee
The latter is composed of two forces: 1- the force due to the gradient of the partial pressure of electron $\dx \pree$ and 2- the Lorentz force.
 The electron heat flux reads :
\begin{multline}
\heate=-\mlambdae\dx \tempe+\left(\pree\mchie+\rhoe h_{\elec}\right)\Ve\\+\pree\sum_{j\in\lourd}\mchiej\Vi[j]+\rhoe h_\elec \sum_{j\in\lourd}\malphaej\Vi[j]
\label{eq:heate}
\end{multline}
where $\mlambdae$ is the electron thermal conductivity tensor and $\rhoe h_\elec $ is the enthalpy of electrons.  $\heate$ is split into two terms : 1- the terms proportional to $\dx\tempe$ and  $\Ve$ are at the heavy particle convective timescale, and 2- the terms proportional to $\Vi,i \in\lourd$ that are at the heavy particle dissipative timescale. 

As in the heavy species transport properties, some usual terms can be identified,i.e, Fick's and Fourier's law. Additionally, terms that couple to the heavy particles diffusion are present at the first order of the generalized Chapman-Enskog expansion. 

In summary, the anisotropic transport coefficients associated to the transport fluxes for electrons are
\begin{center}
	$
	\mDee,\quad \mchie,\quad \mlambdae.
	$
\end{center}

In this section, a list of the transport fluxes and the corresponding transport coefficients has been presented. The method used for computing the latter will be presented in the next section \ref{methodology}.

\subsection{Generalized Ohm's law}%
\label{sec:sec4}
In the following, we derive a general expression for the Ohm's law in the previous set of governing equations \eqref{eq:rhoe}-\eqref{eq:maxwell}. In order to do that, we rewrite the expression of the electric current by grouping the terms in each driving forces. By doing this, we obtain a general algebraic expression for the electric field $\E$ as a function of the transport coefficients and the corresponding driving forces. 

With the electron diffusion velocity (Eq. \eqref{eq:electrondiffusiondrivingforce}) and the heavy particle diffusion velocity (Eq. \eqref{eq:Vi}) we find the total current $\courantel$ as follows
\begin{multline}
	\courantel = \ntot \qtot \vitesse + \frac{\left(\nee\qe\right)^2}{\pree}\matEp\Ep  - \nee\qe \Bigg[\matpee\frac{\dx \pree}{\pree} \\+ \sum_{j\in\lourd}\matpj\frac{\dx\prj}{\prh} + \matTe\glogTe + \matTh\glogTh \Bigg] 
\label{eq:currentTermsNormalized}
\end{multline}
where the multicomponent electromagnetic matrices $\symbolmat$ are defined as:
\begin{align}
\matEp&= \frac{\pree}{\prh}\left[\sumi\mxei\left(\sumi[j]\Dij\mxej\right) + \mDee\right] \label{eq:electromagneti1}, \\
\matpee  &= \frac{\pree}{\prh}\left[\sumi[i]\mxei\left(\sumi[j]\Dij\malphaei[j]\right) + \mDee\right], \label{eq:electromagneti2}\\
\matpj&= \sumi[i]\mxei\Dij,\quad j\in\lourd,\\
\matTe &= \frac{\pree}{\prh}\left[\sumi\mxei\left(\sumi[j]\Dij\mchiej\right) + \mDee\mchie\right], \label{eq:electromagneti3}\\
\matTh&= \left[\sumi\mxei\left(\sumi[j]\Dij\chihj\right)\right],
\label{eq:electromagneti4}
\end{align}
and the tensor $\mxei$ is defined as
\be
\mxei= \frac{\ni\qi}{\nee \qe}\identite + \malphaei, \quad i \in\lourd.
\ee

Using Ampere's law, the general expression of the electric field is obtained
\begin{multline}
	\Ep =\matEp^{-1}\Bigg[ \frac{\pree}{(\nee\qe)^2}\left(\JJe +\JJi[\heavy] \right) + \frac{\pree}{\nee\qe} \Big(\matpee \frac{\dx \pree}{\pree} + \sum_{j\in\lourd}\matpj\frac{\dx\prj}{\prh} \\+\matTe\glogTe + \matTh \glogTh  \Big)\Bigg]
	\label{eq:ElectricField}
\end{multline}
The expression of the multicomponent electromagnetic matrices $\matEp, \matpee, \matTe, \matTh, \matpj j\in\lourd$ can be subdivided into two terms: 1) a term which depend on the coupled heavy particle-electron transport properties, such as $\malphaei[j],\mchiej,\Dij, \chihj \quad i,j \in\lourd $, which scales at the dissipative timescale for the heavy particles at order $\epsilon$, and 2) a term which depend only on the electron transport properties $\mDee,\mchie$, which scales at the convective timescale for the heavy particles at order $\epsilon^0$. 

Some usual terms can be identified in the general expression of the electric field Eq.~\eqref{eq:ElectricField}. The first term of Eq.~\eqref{eq:ElectricField} is the resistive term, where the expression of the resistivity tensor is defined as
\be
 \metae=\frac{\pree}{(\nee\qe)^2}\matEp^{-1}.
\ee
The second term and third term of Eq.~\eqref{eq:ElectricField}, can be identified as a general expression of the battery term for a multicomponent plasma due to the pressure gradients of electrons and heavy particles. The fourth and last term of Eq.~\eqref{eq:ElectricField} are additional terms due to the presence of Soret/Dufour terms in the equations of the diffusion velocities Eq.~\eqref{eq:Vi} and Eq.~\eqref{eq:electrondiffusiondrivingforce}.

In Appendix B, a simplified fully ionized plasma case has been considered which leads to a simplified expression for the expression of the electric field. In this case, the  multicomponent electromagnetic matrices can be simplified, and the usual expression of the electric field and magnetic induction equation are retrieved (see in Appendix B). \\

\section{Methodology}
\label{methodology}
For the purpose of the work, a Helium-Hydrogen mixture, composed of 92\% of Hydrogen and 8\% of Helium which is typical in the Sun atmosphere \cite{Asplund}, is considered, as follows
 \be 
S_1 =\{ \ce{He},~\ce{He^+},~\ce{H},~\ce{H_2},~\ce{He^{++}},~\ce{H^+},~e-\}. \label{eq:s1}
\ee
The heavy species such as carbon, oxygen or metals are not considered.  We assume that they do not impact the transport properties as they are trace elements, i.e., the mole fractions are very small.

We study the transport coefficients for the previous mixture within a range of temperature, pressure, and magnetic field that are typical of the chromosphere \cite{Vernazza,carlsson}:  the temperature varies from $1000$ K to $30000$ K, the pressure from $1$ Pa to $10^4$ Pa, and the magnetic field from a few Gauss to thousands of Gauss \cite{wiegelmann}. In the following, the plasma beta parameter is defined as $\beta_p=2\mu_0\pression/|\B|^2$, where $\pression$ is the total pressure of the plasma in Pascal and $|\B|$ is the magnetic field in Tesla.

Consequently, for a range of temperature between $1000$ K and $30 000$ K, two cases have been considered. The \textbf{case A}, where the total pressure is $\pression=10^4$ Pa, and $\beta_p=10$, which represents typical conditions of the bottom of the photosphere. The \textbf{case B}, where the total pressure is $\pression=1$ Pa, and $\beta_p=0.1$, representing a magnetized region of the upper chromosphere.

\begin{figure}[!]
	\centering
	\includegraphics[trim={0cm 0 0cm 0},width=\columnwidth,clip]{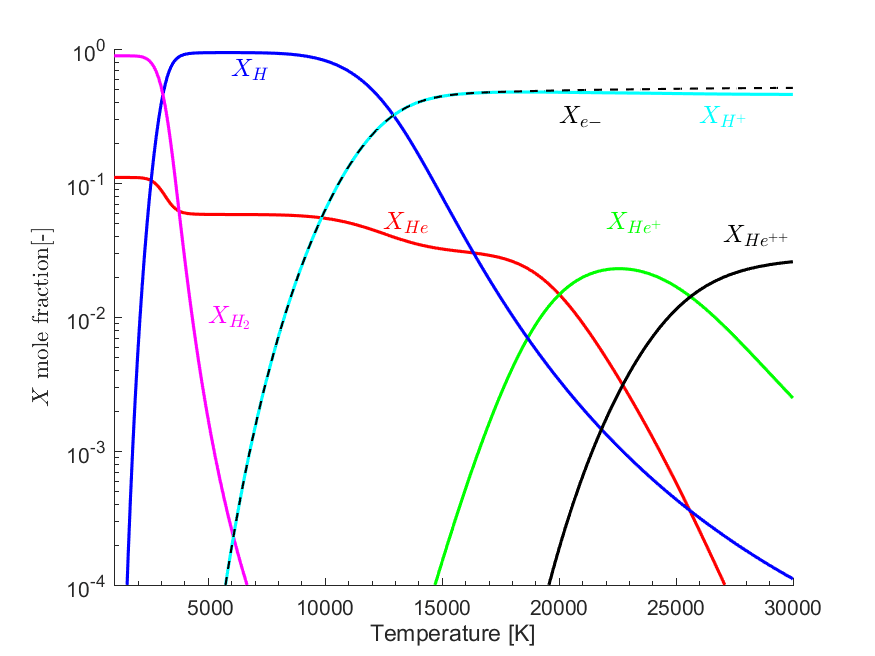}%
	\caption{Mole fraction of the Helium-Hydrogen mixture $S_1$, for  $P=10^{4}$ Pa (\textbf{case A}), as a function of temperature}%
	\label{fig:ionization1}%
\end{figure}

\begin{figure}[!]
	\centering
	\includegraphics[trim={0cm 0 0cm 0},width=\columnwidth,clip]{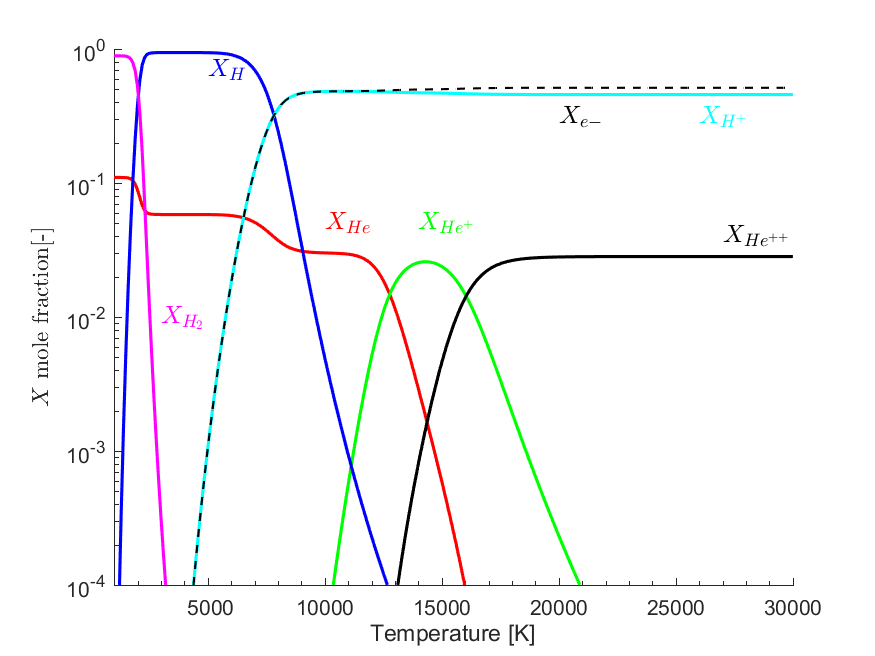}%
	\caption{Mole fraction of the Helium-Hydrogen mixture $S_1$, for $P=1$ Pa (\textbf{case A}), as a function of temperature}%
	\label{fig:ionization2}%
\end{figure}

\begin{figure}[!]
	\centering
	\includegraphics[trim={0cm 0 0cm 0},width=\columnwidth,clip]{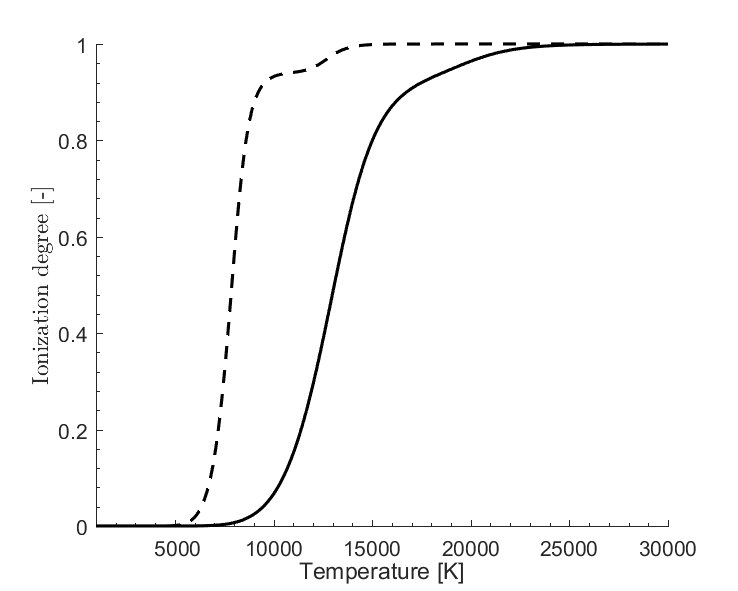}%
	\caption{Ionization degree of $S_1$, for \textbf{case A} and \textbf{case B} as function of the temperature. \textbf{Case A} in full line, \textbf{case B} in dashed line}%
	\label{fig:ionization3}%
\end{figure}

Based on the chosen conditions, we compute the thermochemical equilibrium composition. The mole fraction and the ionization degree of the Helium-Hydrogen mixture $S_1$ for the \textbf{case A} and \textbf{case B} are shown in Fig.~\ref{fig:ionization1}, Fig.~\ref{fig:ionization2} and Fig.~\ref{fig:ionization3}. These results are obtained with a method that is based on the minimization of the Gibbs free energy with suitable mass balance constraints \cite{gibbsfree} in thermal equilibrium. The compositions that are shown in Figs.~\ref{fig:ionization1} and \ref{fig:ionization2} will be used to study the transport properties in the following sections.  


The calculation of the transport coefficients is based on the solution of integro-differential equations. In order to solve these equations, the spectral Galerkin method is applied. This method expands the coefficients in a series of orthogonal Laguerre-Sonine polynomials that are truncated at a given order of approximation. The calculation is thus reduced to a linear algebraic system of equations. As a result, the transport coefficients can be obtained by the resolution of determinants that are known functions of the macroscopic parameters, i.e., the field variables and the collision integrals between particles. The solution of these systems allows for the transport coefficients to be written as linear combinations of the collision integrals, which take into account the interaction potential for a collision between two particles. These linear combinations are derived by extending the definition and the calculation of bracket integrals introduced by Ferziger and Kapper \cite{ferzigerkapper} or in \cite{kinetictheory,zhdanov,balescu} to the thermal nonequilibrium case, studied in depth by Kolesnikov \cite{Kolesnikov}. According to Magin \& Degrez \cite{magin04}, Kolesnikov \cite{Kolesnikov} and Tirsky \cite{Tirsky}, the transport coefficients involving collisions between heavy particles and electrons converge for expansions in second order non vanishing Laguerre-Sonine polynomials and higher. In this work, we use the third order Laguerre Sonine polynomials approximation in order to compute  the transport properties. This method has been widely used in the literature (e.g., \cite{bruno,kinetictheory,Transportproperties,zhdanov,balescu,capitelli}).

The explicit relations for $\malphaej, \mchiej,$ with $j \in \lourd$ and $\mDee,\mchie,\mlambdae$ in terms of the solutions to the transport systems can be found in Scoggins et al.~\cite{scoggins}. The heavy particle transport systems for $\etah, \Dij, \chihj, \lambdaH,$ with $i,j\in\lourd$ are found in Magin \& Degrez \cite{Transportproperties} with the difference that the mole fractions are given in terms of heavy species only, excluding electrons.

\section{Verification of the method in a fully ionized plasma case $S_2$}
\label{sec:sec5}
In order to verify the presented method, we perform a comparison with Braginskii's transport properties. In Braginskii \cite{Braginskii}, the method that is used for the computation of the transport properties as well as the derivation of the governing equations are identified and valid only for fully ionized plasmas. The objective is to validate the method that is used for the computation of the transport properties.

As it can be seen in Fig.~\ref{fig:ionization3}, the Helium-Hydrogen mixture $S_1$ can be considered to be fully-ionized, mainly composed of $S_2 = \{\ce{H^+},\ce{e-}\}$, when the temperature is higher than $15000$ K. The comparison will be thus performed in conditions where the mixture is $S_2$ in a range of temperatures from $T=15000$ K to $T=30000$ K for the \textbf{case A} and \textbf{case B}. To illustrate the comparison, we focus on the properties $\lambdaepa$, $\lambdaepe$, $\etah$ and $\lambdaH$, although the rest of preperties show similar behaviour.

On the one hand, in \cite{Braginskii}, the derivation of the governing equations can be summarized in three main steps: 1- A fully ionized ion-electron plasma is considered in a constant magnetic field, 2- The Landau collision operators are used, simplified by the Lorentz process, and 3- an adapted Chapman-Enskog method is used based on the square root of the mass ratio between electron and ions \cite{balescu}. On the other hand, in Graille et al.~ \cite{graille}, a general multicomponent plasma that can be partially or fully ionized is considered in a constant magnetic field, the Chapman and Cowling collision operators highly studied in \cite{ferzigerkapper,kinetictheory} are used and the Chapman-Enskog expansion is performed after a non-dimensional analysis of the Boltzmann equation. Finally, the two methods lead to distinct governing equations. 

Although the governing equations between the two models are different, the integro-differential systems for computing the transport properties are similar or even identical in the case of a fully ionized plasma. As a matter of fact, in both models, the anisotropic electron transport properties have the same integro-differential systems. However, only the integro-differential systems related to the parallel component of the heavy particle transport properties are identical to those from the model derived by Graille et al.~\cite{graille}. Consequently, only the parallel component of the heavy particle transport properties can be compared with those from the model of Graille et al~\cite{graille}. This is due to the fact that both models are based on the Chapman-Enskog expansion. However, the differences result from the scale analysis from the Boltzmann equation that is carried out by Graille et al.~\cite{graille} before applying the expansion.

In both models, the transport coefficients are expanded in a series of orthogonal Laguerre-Sonine polynomials. The latter are written as linear combinations of collision integrals that are simplified by potential interactions, based on the usual Coulomb interaction cut-off at the Debye-length. This approximation assumes collisions with large impact parameters and small scattering angles. However, in Braginskii \cite{Braginskii}, the series are truncated at the second order approximation \cite{balescu} whereas a third order approximation has been performed in the case of the presented model. The expression of the transport coefficients depends on the mean collision times $\taueB$ and $\tauhB$ defined as
\be 
\taueB=\frac{3 {\refme}^{2} {\epsilon_0}^{2} }{\nH \qe^4 \log(\Lambda)}
\left(\frac{2 \pi \boltz \tempe}{\qe^2 \nee}\right)^{\frac{3}{2}}, \quad
\tauhB=\sqrt{\frac{2 \refmh}{\refme}} \left(\frac{\temph}{\tempe}\right)^{\frac{3}{2}}Z^{-2} \taueB,
\label{eq:tau} 
\ee
where $\log(\Lambda)$ is the Coulomb logarithm defined by Spitzer \cite{Spitzer}, and $Z$ is the charge number. The mean collision times as defined in Eq.\eqref{eq:tau}, can be seen as a first order Chapman Cowling approximation of the collision time for electron/ion and ion/ion collisions \cite{kinetictheory}. Correction terms depending on $Z$ are used for the computation of the transport coefficients. This method leads to simplified expressions of the transport coefficients that depend only on the mean collision times and the charge number of the fully ionized plasma considered \cite{balescu,kinetictheory}.  


In Braginskii \cite{Braginskii}, the parallel and perpendicular components of the electron thermal conductivity tensor are defined as  
\be
\lambdaepa\underset{\textbf{Br}}{\;:=\;}\frac{\nee \boltz^{2}\tempe }{\refme}\taueB\left[3.16\right],
\label{eq:lambdaepabr}
\ee
\be
\lambdaepe\underset{\textbf{Br}}{\;:=\;}\frac{\nee \boltz^{2}\tempe }{\refme}\taueB \left[\frac{4.664 x^2+11.92}{x^4+14.79 x^2+3.77}\right],
\label{eq:lambdaepebr}
\ee
where $\textbf{Br}$ denotes the computation of the transport coefficient as dervied by Braginskii \cite{Braginskii}.  $x=\omega_\elec \taueB$ and $\omega_\elec=\qe \B/\refme$ and the values in brackets correspond to Braginskii's coefficients for a charge number $Z=1$.

\begin{figure}[!]
	\centering
	\includegraphics[trim={0cm 0 0cm 0},width=\columnwidth,clip]{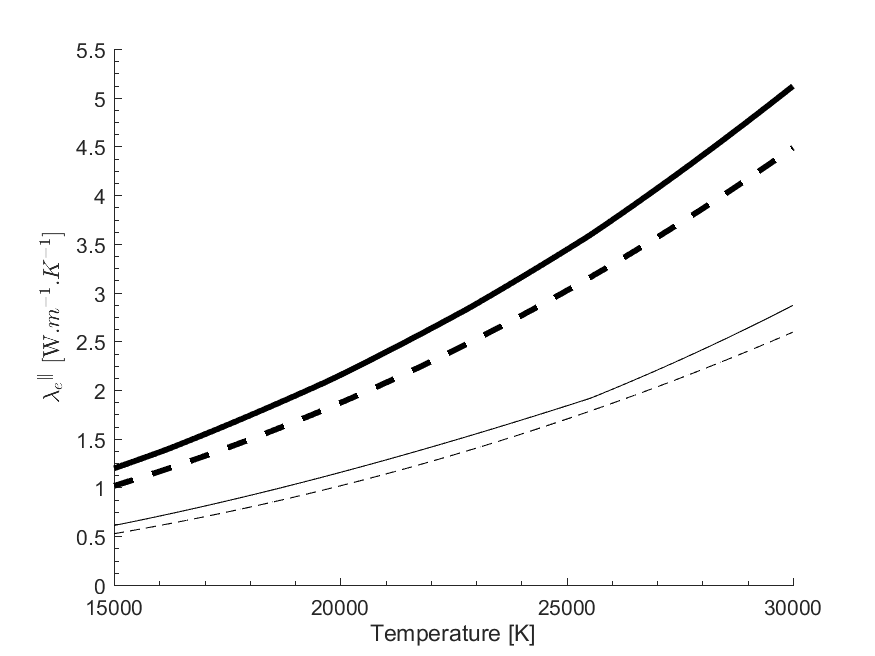}%
	\caption{Parallel component of the electron thermal conductivity tensor $\lambdaepa$ for a fully ionized plasma $S_2$, as function of temperature: Dashed lines and full lines correspond to the transport coefficient from the model of Braginskii \cite{Braginskii}, and from Graille et al.~\cite{graille} respectively. Bold lines correspond to the \textbf{case A}, the other lines correspond to the \textbf{case B}}%
	\label{fig:tauelectronthermalconduc}%
\end{figure}

\begin{figure}[!]
	\centering
	\includegraphics[trim={0cm 0 0cm 0},width=\columnwidth,clip]{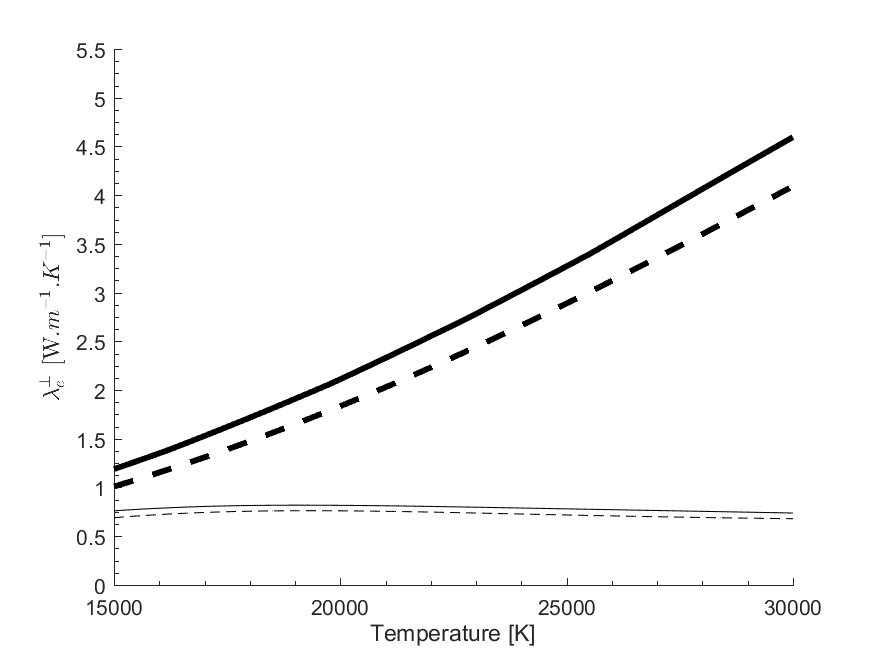}%
	\caption{Perpendicular component of the electron thermal conductivity tensor $\lambdaepe$ for a fully ionized plasma $S_2$, as function of temperature: Dashed lines and full lines correspond to the transport coefficient from the model of Braginskii \cite{Braginskii}, and from  Graille et al.~\cite{graille} respectively. Bold lines correspond to the \textbf{case A}, the other lines correspond to the \textbf{case B}}%
	\label{fig:tauelectronthermalconduc2}%
\end{figure}

Fig.~\ref{fig:tauelectronthermalconduc} and Fig.~\ref{fig:tauelectronthermalconduc2} show the parallel and perpendicular component of the electron thermal conductivity tensor $\mlambdae$, as function of the temperature, for the \textbf{case A} and the \textbf{case B}, for the fully ionized plasma $S_2$. Here, we compare the expressions from Braginskii Eq.~\eqref{eq:lambdaepabr} and Eq.~\eqref{eq:lambdaepebr} with those that are given by Scoggins \cite{scoggins} that are based on a third order Laguerre-Sonine polynomials approximation.

Strong similarities are obtained in all the considered cases. In ~Braginskii \cite{Braginskii}, the components of the electron thermal conductivity tensor are underestimated leading to differences that are less than $20\%$. These differences are increasing at high temperatures. Similar results have been obtained for all the other electron transport properties.


Similarly, the parallel component of the heavy thermal conductivity and of the heavy particle viscosity of the model of Braginskii~\cite{Braginskii}, have been compared with the expression from \cite{scoggins} and \cite{magin04}. In Braginksii \cite{Braginskii}, the heavy thermal conductivity and heavy particle viscosity are defined as
\be
\lambdaH^{\parallel}\underset{\textbf{Br}}{\;:=\;}\nH\boltz^{2}\temph \tauhB\left[ 3.91 \right],
\label{eq:lambdah}
\ee
\be
\etah^{\parallel}\underset{\textbf{Br}}{\;:=\;} \nH \boltz \temph\tauhB\left[0.96\right].
\label{eq:etah}
\ee

\begin{figure}[!]
	\centering
	\includegraphics[trim={0cm 0 0cm 0},width=\columnwidth,clip]{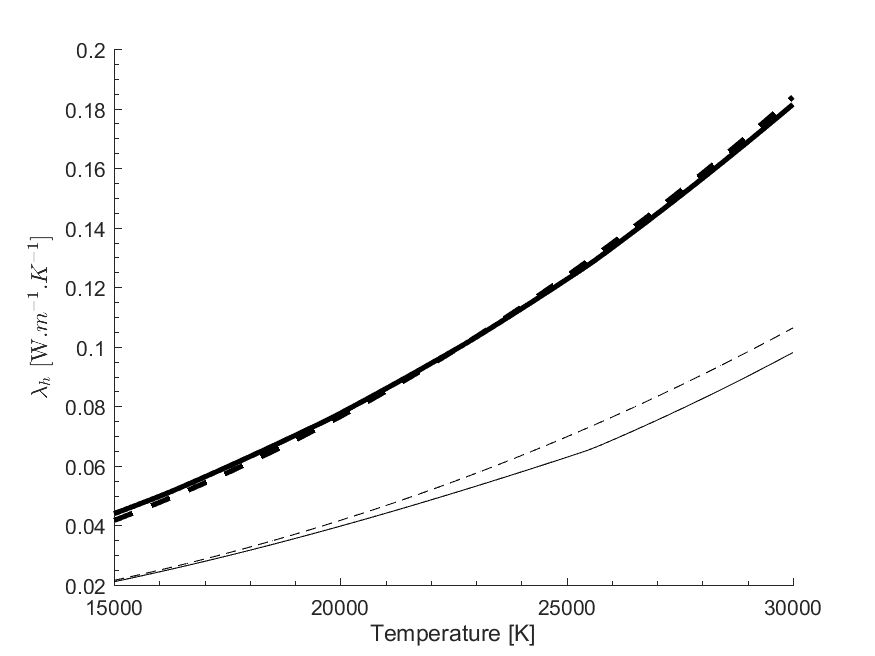}%
	\caption{Heavy thermal conductivity  $\lambdaH$ for a fully ionized plasma $S_2$, as function of temperature: Dashed lines and full lines correspond to the transport coefficient from the model of Braginskii\cite{Braginskii}, and from  Graille et al.~ \cite{graille} respectively. Bold lines correspond to the \textbf{case A}, the other lines correspond to the \textbf{case B}}%
	\label{fig:lambdahhh}%
\end{figure}

\begin{figure}[!]
	\centering
	\includegraphics[trim={0cm 0 0cm 0},width=\columnwidth,clip]{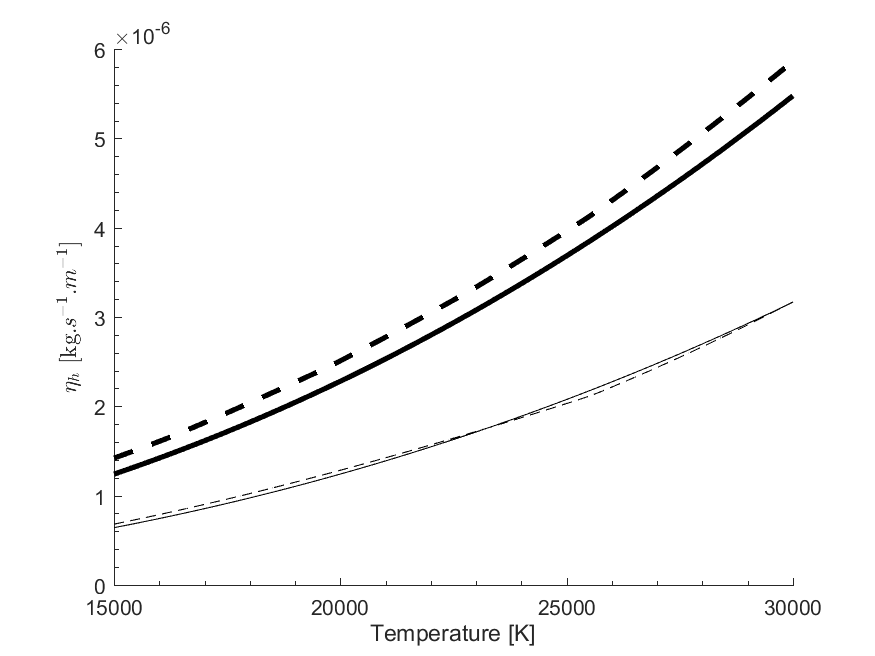}%
	\caption{ Heavy particle viscosity $\etah$ for a fully ionized plasma $S_2$, as function of temperature: Dashed lines and full lines correspond to the transport coefficient from the model of Braginskii\cite{Braginskii}, and from  Graille et al.~\cite{graille} respectively. Bold lines correspond to the \textbf{case A}, the other lines correspond to the \textbf{case B}}%
	\label{fig:etah}%
\end{figure}

Fig.~\ref{fig:lambdahhh} and Fig.~\ref{fig:etah} show the heavy thermal conductivity $\lambdaH$ and the heavy particle viscosity $\etah$ respectively, in the same conditions as in Fig.~\ref{fig:tauelectronthermalconduc} and Fig.~\ref{fig:tauelectronthermalconduc2}.

As it was seen before, strong similarities have been obtained in all the considered cases for the chosen conditions, which leads to differences that are smaller than $20\%$. In addition, it can be shown that the heavy transport properties from \cite{Braginskii} are isotropic at the chosen conditions.

In summary, we can conclude that the proposed method is verified for the fully ionized case. The main differences that are obtained between the two models are due to 1-the order of Laguerre Sonine polynomials that was used, i.e., second order in Braginskii's model \cite{balescu} and third order in the proposed method, and 2- the nature of the collision operators used, Landau collision operators in the model of Braginskii and Chapman and Cowling collision operators in the model of Graille et al.~\cite{graille}. Additionally, the formulation of the transport properties that are considered in this paper are generalized for any type of partially ionized mixture.


\section{Transport properties for a partially ionized Helium-Hydrogen plasma} 
\label{sec:sec3}

\subsection{Transport fluxes in thermo-chemical equilibrium}

In order to simplify the analysis of the presented transport systems, we consider thermochemical equilibrium  $\tempe=\temph=T$, isobaric mixtures at rest. The total heat flux is entirely a function of the temperature gradient and magnetic field and may be written as 
\be
\heati[\heavy]+\heate = -\left(\lambdaH +\mlambdae + \mlambdaS+\mlambdaR\right)\dx T,
\label{eq:heatflux_simple}
\ee
where the Soret and reactive thermal conductivities may be written as 
\be
\mlambdaS = -\pree \mchie \theta_{\elec} -\sum_{j\in\lourd}\left[\prh\chihj +\pree\mchie\right] \theta_{i},
\label{eq:soretthermalconductivity}
\ee 
\be
\mlambdaR = -\rhoe h_\elec \theta_{\elec} -\sum_{j\in\lourd}\left[\rhoi[j]h_{j} +\rhoe h_\elec \malphaej \right]\theta_{j}.
\label{eq:reactivethermalconductivity}
\ee 
where $\theta_{\elec}$ and $\theta_{i}, i\in\lourd$ are defined as
\be
\theta_{\elec} = -\mDee\left[\frac{1}{x_{\elec}}\frac{\partial x_{\elec}}{\partial T}+\frac{\mchie}{T} \right],
\ee
\be
\theta_{i} = \sum_{j\in\lourd}\Dij\left[\frac{1}{1-x_{\elec}}\left(\frac{\partial x_{i}}{\partial T}+\frac{\partial x_{\elec}}{\partial T}\malphaej \right) + \frac{\chihj}{T} + \frac{\pree}{\prh}\frac{\mchiej}{T}\right], i \in \lourd.
\ee
Where $\theta_{\elec}$ and $\theta_{i}, i\in\lourd$ correspond to diffusion velocities for a temperature gradient of 1, i.e, $\Ve=\theta_{\elec}\dx T$ and $\Vi=\theta_i \dx T$.
Using the Mutation++ library \cite{mutation}, we compute all the transport properties for the Helium-Hydrogen mixture $S_1$ for \textbf{case A}, i.e., weakly magnetized, and \textbf{case B} (magnetized case).

\begin{figure}[!]
	\centering
	\includegraphics[trim={0cm 0 0cm 0},width=\columnwidth,clip]{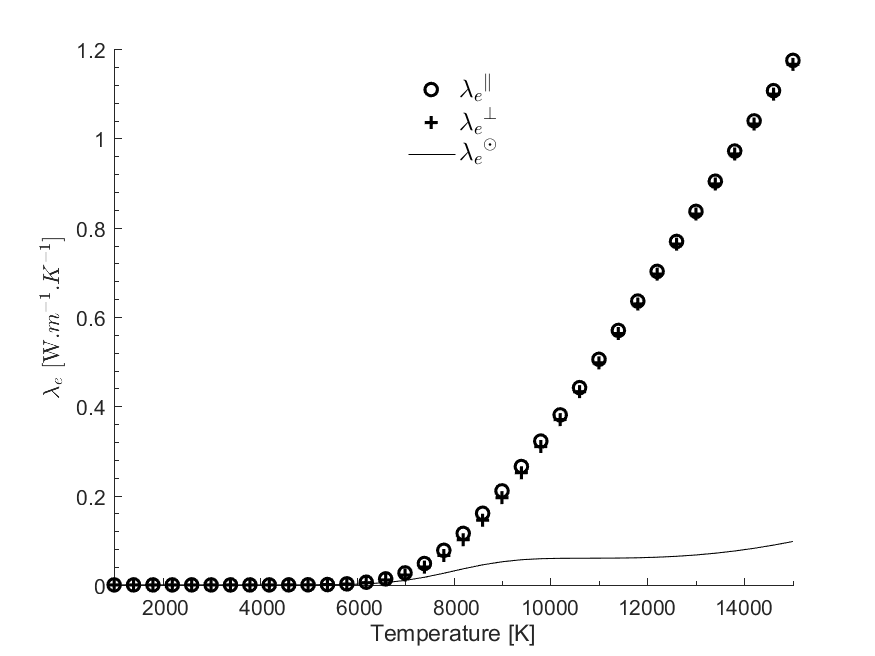}%
	\caption{Components of the electron thermal conductivity tensor $\mlambdae$ for the isotropic \textbf{case A}, at the third order Laguerre Sonine polynomials, for the Helium-Hydrogen mixture $S_1$ as a function of temperature}%
	\label{fig:partiallytransport2}%
\end{figure}

\begin{figure}[!]
	\centering
	\includegraphics[trim={0cm 0 0cm 0},width=\columnwidth,clip]{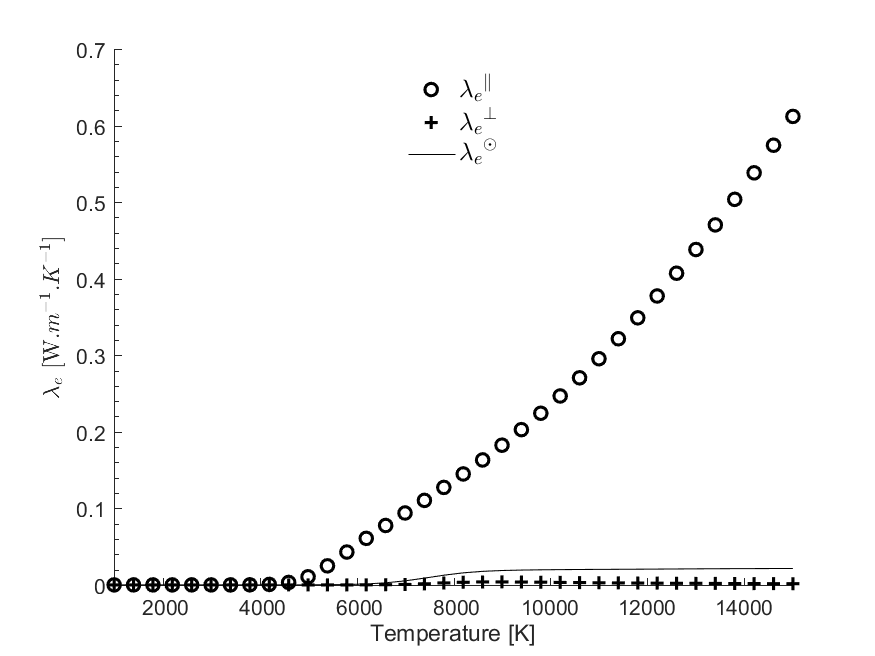}%
	\caption{Components of the electron thermal conductivity tensor $\mlambdae$ for the anisotropic \textbf{case B}, at the third order Laguerre Sonine polynomials, for the Helium-Hydrogen mixture $S_1$ as a function of the temperature}%
	\label{fig:partiallytransport1}%
\end{figure}

Fig.~\ref{fig:partiallytransport2} and Fig.~\ref{fig:partiallytransport1} present the parallel, perpendicular and transverse components of the electron thermal conductivity tensor $\mlambdae$ as a function of the temperature, for both cases.
According to Fig. \ref{fig:partiallytransport2} (\textbf{case A}), the perpendicular component is equal to the parallel component for the entire range of temperatures, i.e, the electron thermal conductivity is isotropic. Indeed, the pressure forces are dominating the magnetic pressure forces, so the plasma is unmagnetized. On the other hand, in Fig. \ref{fig:partiallytransport1} (\textbf{case B}), for higher temperatures than $T=5000$ K, the electron thermal conductivity $\mlambdae$ is anisotropic since the magnitude of magnetic field is higher. This results in a transverse component that is higher than the perpendicular component of $\mlambdae$. Similar results have been obtained for the other anisotropic electron transport properties such as $\mDee$ and $\mchie$.

\begin{figure}[!]
	\centering
	\includegraphics[trim={0cm 0 0cm 0},width=\columnwidth,clip]{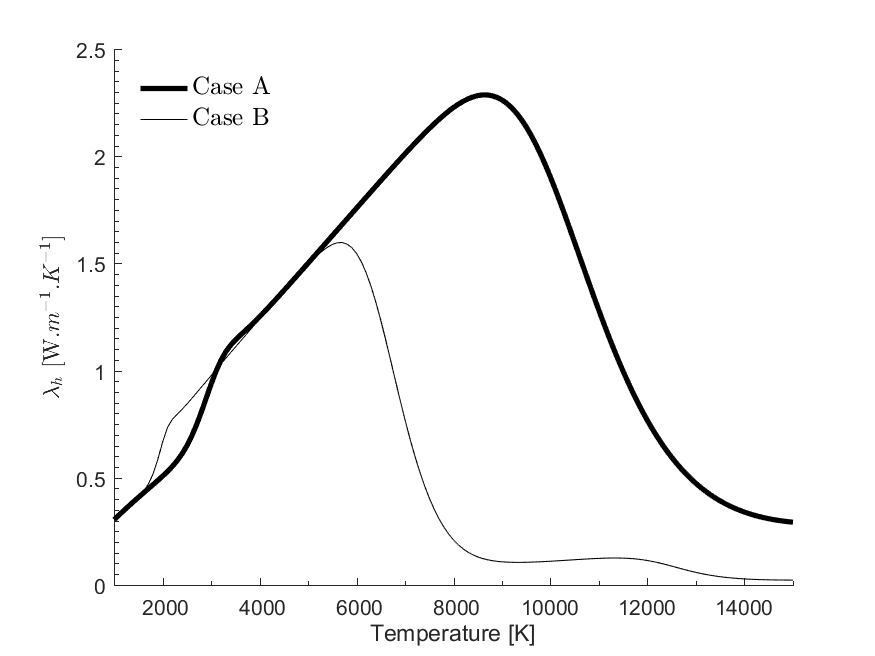}%
	\caption{Heavy thermal conductivity $\lambdaH$, at the third order of Laguerre Sonine approximation, for \textbf{case A} and \textbf{case B}, for the Helium-Hydrogen mixture $S_1$ as a function of temperature }%
	\label{fig:lambdahtransport1}%
\end{figure}

Fig. \ref{fig:lambdahtransport1} shows the heavy particle thermal conductivity $\lambdaH$ as a function of the temperature, for the \textbf{case A} and \textbf{case B}.
In Fig. \ref{fig:lambdahtransport1}, strong differences between the two cases for a temperature higher than $6000$ K can be seen. 
In the \textbf{case A}, $\lambdaH$ increases from $1000$ K to $9000$ K, which is expected since $\lambdaH$ is an increasing function of the temperature. However, in the \textbf{case A} after $9000$ K, $\lambdaH$ decreases. This decrease is due to the ionization of the hydrogen. Indeed, the heavy particle thermal conductivity is related to a combination of the cross sections variations of all the heavy species in the mixture, which are proportional to the mole fractions of each heavy particles. This result is coherent with Fig. \ref{fig:ionization1}, which shows that the mole fraction of $\ce{H}$ is decreasing after $9000$ K.
Similar behavior as the  \textbf{case A} have been observed for the \textbf{case B}, except that the ionization of \ce{H} starts at $6000$ K for this pressure. In Fig. \ref{fig:lambdahtransport1}, the second modulation observed around $12000$ K is due to the ionization of the Helium as shown in Fig. \ref{fig:ionization2}.

\begin{figure}[!]
	\centering
	\includegraphics[trim={0cm 0 0cm 0},width=\columnwidth,clip]{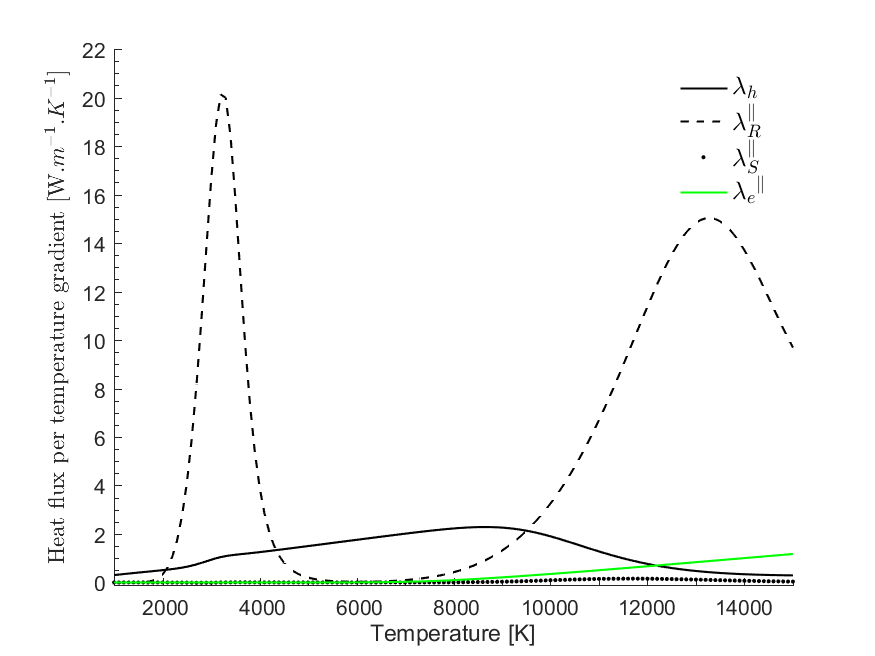}%
	\caption{Component of the total heat flux (\ref{eq:heatflux_simple}) as a function of the temperature for the isotropic \textbf{case A} for the Helium-Hydrogen mixture $S_1$}%
	\label{fig:lambdacompa_hp}%
\end{figure}

Fig. \ref{fig:lambdacompa_hp} shows the components of the total heat flux (Eq.~\ref{eq:heatflux_simple}) as a function of the temperature, for the isotropic \textbf{case A}. It is clear that the reactive thermal conductivity $\lambdaRpa$ is higher than the other components for certain ranges of temperature between $2200$ K and $4300$ K  and for temperature higher than $10000$ K. The heavy thermal conductivity $\lambdaH$ is the second term which dominates the total heat flux, and is higher than $\lambdaRpa$ for a range of temperature between $4200$ K and $10 000$ K. The results here obtained are consistent with those of Scoggins et al.~\cite{scoggins}.

\begin{figure}[!]
	\centering
	\includegraphics[trim={0cm 0 0cm 0},width=\columnwidth,clip]{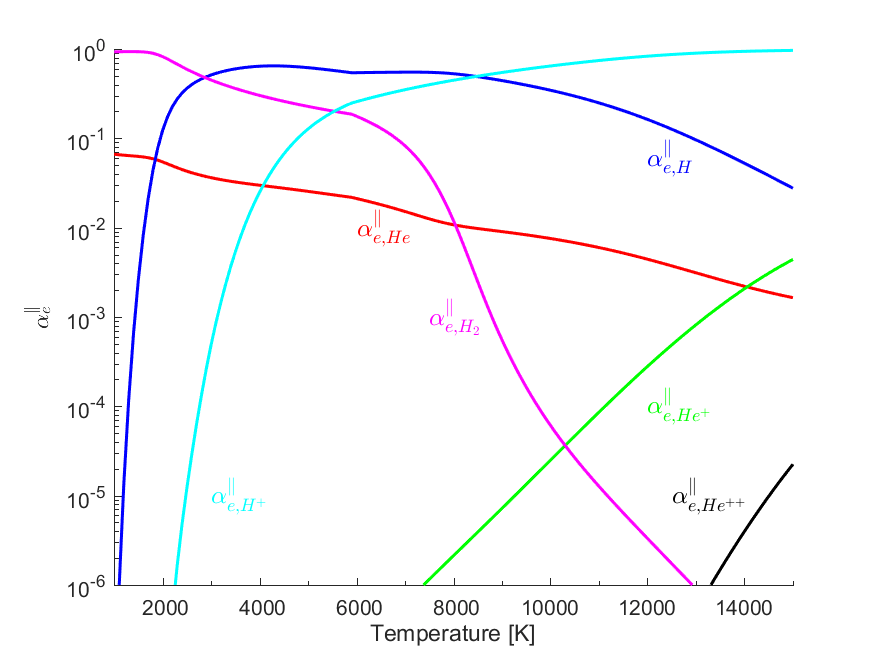}%
	\caption{Parallel component of the electron-heavy particle  transport coefficient $\alphaeipa[j], j\in\{ \ce{He}, \ce{He^+}, \ce{H}, \ce{H_2},\ce{He^{++}},\ce{H^+}\}$, for the isotropic \textbf{case A}, for the Helium-Hydrogen mixture $S_1$}%
	\label{fig:kolesnikov1}%
\end{figure}

\begin{figure}[!]
	\centering
	\includegraphics[trim={0cm 0 0cm 0},width=\columnwidth,clip]{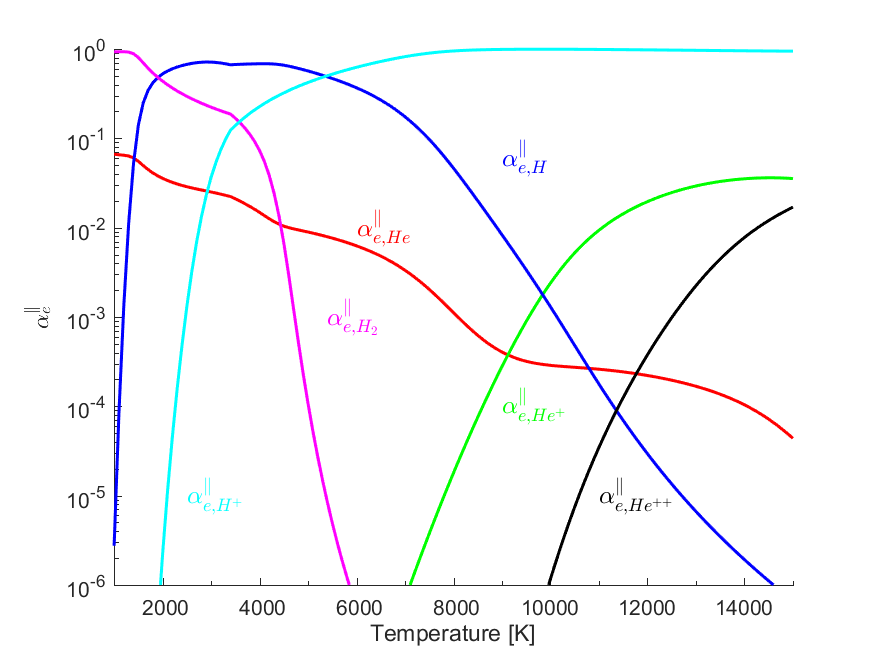}%
	\caption{Parallel component of the electron-heavy particle  transport coefficient $\alphaeipa[j], j\in\{ \ce{He}, \ce{He^+}, \ce{H}, \ce{H_2},\ce{He^{++}},\ce{H^+}\}$, for the anisotropic \textbf{case B}, for the Helium-Hydrogen mixture $S_1$}%
	\label{fig:kolesnikov2}%
\end{figure}

Fig. \ref{fig:kolesnikov1} and Fig. \ref{fig:kolesnikov2} show the parallel component of each term of the electron-heavy particle transport coefficients $\alphaeipa[j], j\in\{ \ce{He}, \ce{He^+}, \ce{H}, \ce{H_2},\ce{He^{++}},\ce{H^+}\}$, as a function of the temperature, for the \textbf{case A} and \textbf{case B}. As before, each term of the electron-heavy particle transport tensor $\alphaeipa[j]$, is proportional to the mole fraction. 


\subsection{Transport properties in a pore in the low Sun chromosphere}	
\label{sec:pore}
As done in the previous section, the transport coefficients of the previous Helium-Hydrogen mixture are computed for the conditions found in the upper layer of the solar convective zone from the radiative 3D MHD simulations of a pore by Kitiashvili et al. \cite{Irina2}. The simulation results are obtained for the computational domain of $6.4\times6.4\times5.5$ Mm with the grid sizes: $50 \times 50 \times 43$ km, $25 \times 25 \times 21.7$ km and $12.5 \times 12.5 \times 11$ km ($1282 \times 127$, $2562 \times 253$ and $5122 \times 505$ mesh points). The domain includes a top, 5 Mm-deep, layer of the convective zone and the low Sun chromosphere. A thermal equilibrium case $\tempe=\temph$ has been considered for the computation of the transport coefficients. 

\begin{figure}[!]
	\centering
	\includegraphics[trim={0cm 0 0cm 0},width=\columnwidth,clip]{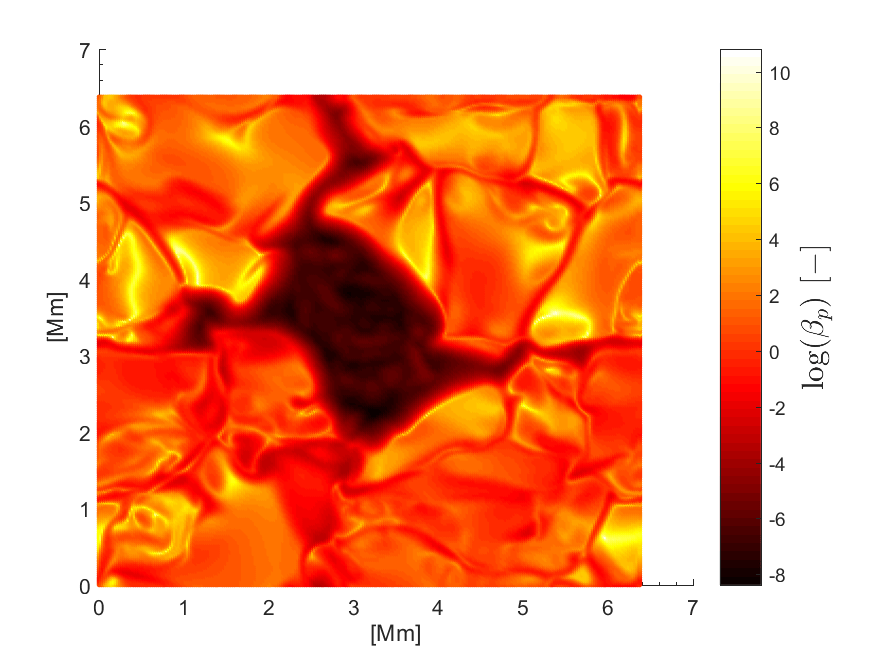}%
	\caption{Plasma beta coefficient $\beta_p$ distribution from the radiative 3D MHD simulations of a pore by Kitiashvili et al.\cite{Irina2}} %
	\label{fig:betappore}%
\end{figure}

\begin{figure}[!]
	\centering
	\includegraphics[trim={0cm 0 0cm 0},width=\columnwidth,clip]{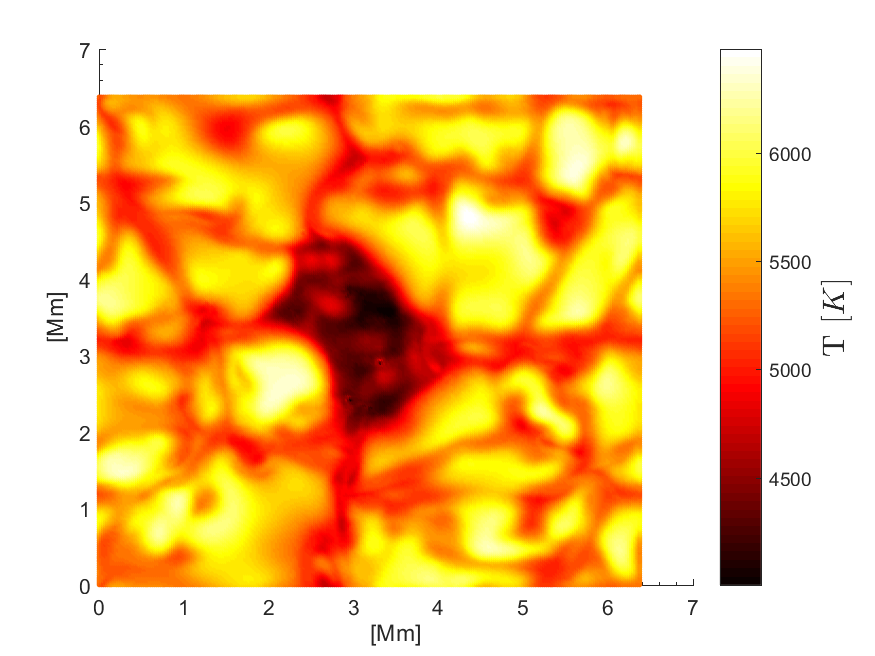}%
	\caption{Temperature (K) distribution from the radiative 3D MHD simulations of a pore by Kitiashvili et al.\cite{Irina2}} %
	\label{fig:Tpore}%
\end{figure}

\begin{figure}[!]
	\centering
	\includegraphics[trim={0cm 0 0cm 0},width=\columnwidth,clip]{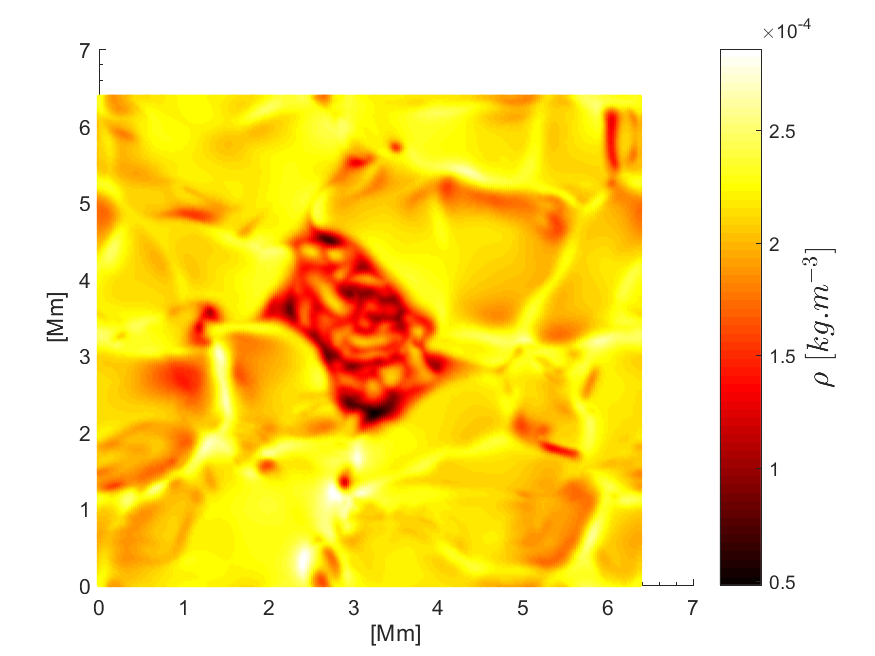}%
	\caption{Total density ($kg.m^{-3}$) distribution from the radiative 3D MHD simulations of a pore by Kitiashvili et al.\cite{Irina2}} %
	\label{fig:rhopore}%
\end{figure}

Figs.\ref{fig:betappore}, \ref{fig:Tpore}, and Fig.(\ref{fig:rhopore}) show snapshots of the distribution of the plasma beta parameter $\beta_p$, temperature $T$, and total density, respectively. As it can be seen, the temperature is varying from $4000$ K to $6500$ K, the plasma beta parameter is varying on a large range of magnitude, from weakly- to strongly-magnetized. In the snapshot of the simulation, a characteristic granulation pattern with the relatively hot ($T > 5500$ K) and less dense upflowing weakly-magnetized plasma in the middle of the granular cells can be observed. In addition, the lower temperature ($T < 4500$ K) and higher density downflowing strongly-magnetized plasma at the intergranulation boundaries can be perceived (red lines of granulation). A strongly-magnetized cold plasma can be seen in the middle of the snapshot. 

\begin{figure}[!]
	\centering
	\includegraphics[trim={0cm 0 0cm 0},width=\columnwidth,clip]{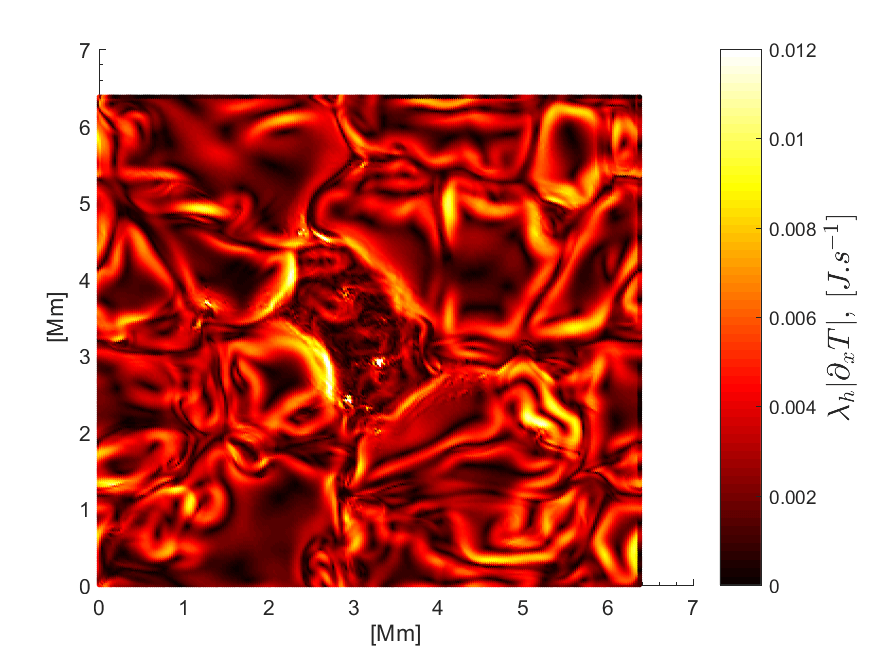}%
	\caption{Distribution of the heavy particle heat flux $\lambdaH |\dx T|$ , for the Helium-Hydrogen mixture $S_1$  based on the results of the  radiative 3D MHD simulations of a pore by Kitiashvili et al.\cite{Irina2}} %
	\label{fig:lambdahpore}%
\end{figure}

\begin{figure}[!]
	\centering
	\includegraphics[trim={0cm 0 0cm 0},width=\columnwidth,clip]{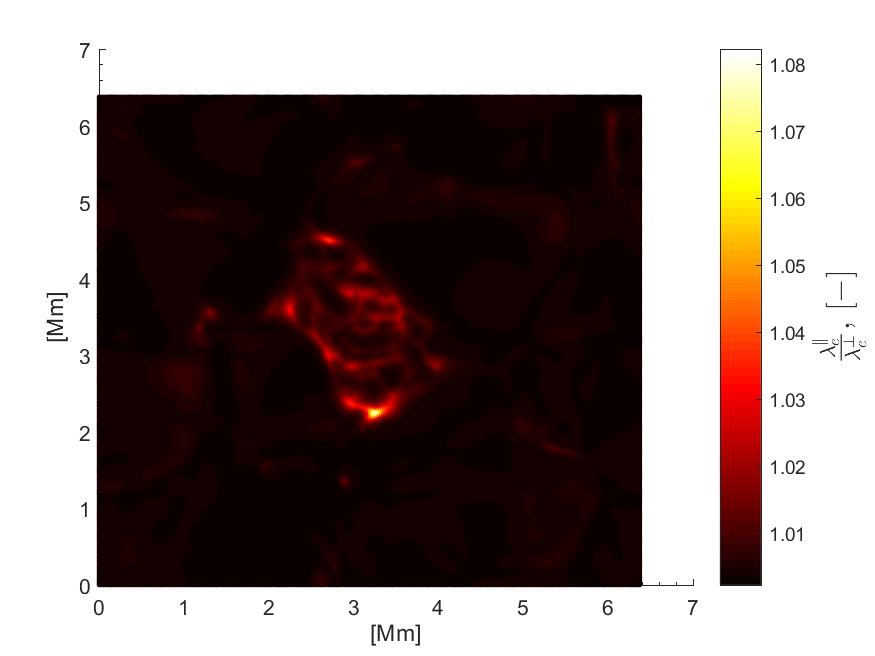}%
	\caption{Ratio $\lambdaepa/\lambdaepe$ distribution, computed at the third order of the Laguerre-Sonine polynomials approximation, for the Helium-Hydrogen mixture $S_1$  based on the results of the  radiative 3D MHD simulations of a pore by Kitiashvili et al.\cite{Irina2}} %
	\label{fig:ratiolambdae}%
\end{figure}

\begin{figure}[!]
	\centering
	\includegraphics[trim={0cm 0 0cm 0},width=\columnwidth,clip]{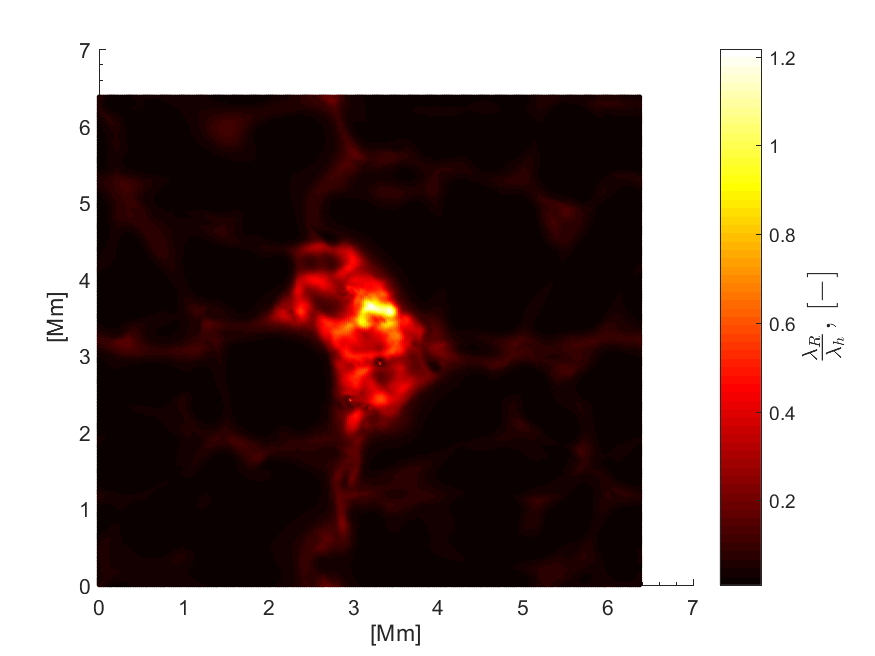}%
	\caption{Ratio $\lambdaRpa/\lambdaH$ distribution, computed at the third order of the Laguerre-Sonine polynomials approximation, for the Helium-Hydrogen mixture $S_1$  based on the results of the  radiative 3D MHD simulations of a pore by Kitiashvili et al.\cite{Irina2}} %
	\label{fig:ratiolambdaRlambdah}%
\end{figure}

\begin{figure}[!]
	\centering
	\includegraphics[trim={0cm 0 0cm 0},width=\columnwidth,clip]{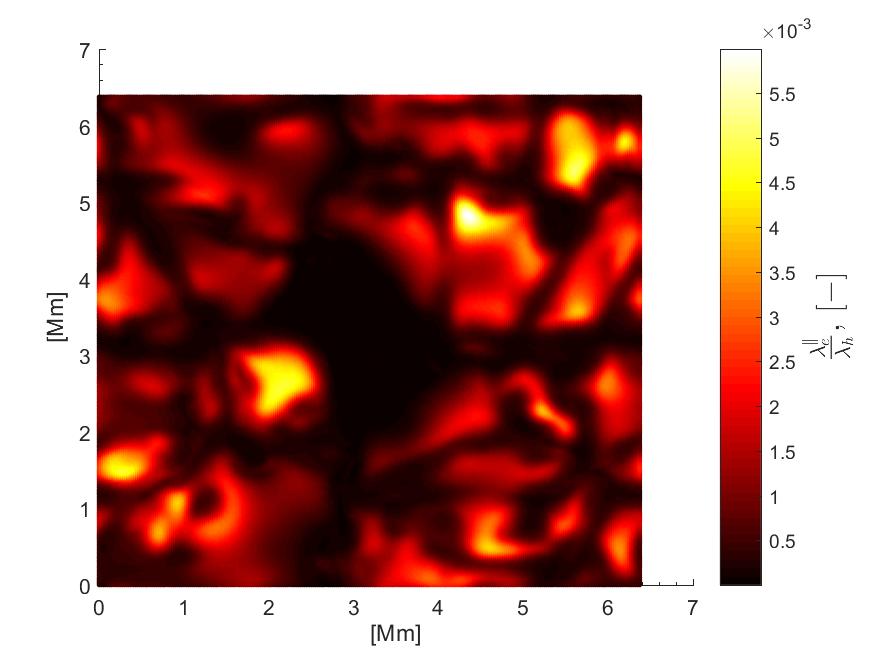}%
	\caption{Ratio $\lambdaepa/\lambdaH$ distribution, computed at the third order of the Laguerre-Sonine polynomials approximation, for the Helium-Hydrogen mixture $S_2$  based on the results of the  radiative 3D MHD simulations of a pore by Kitiashvili et al.\cite{Irina2}} %
	\label{fig:ratiolambdaelambdah}%
\end{figure}

Figs. \ref{fig:lambdahpore}, \ref{fig:ratiolambdae}, \ref{fig:ratiolambdaRlambdah}, and \ref{fig:ratiolambdaelambdah} present the distribution of the heavy particle heat flux $\lambdaH|\dx T|$, the ratio $\lambdaepa/\lambdaepe$, $\lambdaRpa/\lambdaH$, and $\lambdaepa/\lambdaH$, respectively. Fig. \ref{fig:ratiolambdae} shows that the electron thermal conductivity tensor $\mlambdae$ is almost isotropic everywhere, except in the middle of the snapshot where $\lambdaepa/\lambdaepe=1.08$. Fig. \ref{fig:ratiolambdaelambdah} shows that the electron thermal conductivity is small compared to the heavy thermal conductivity. This results is related to the results from Fig. \ref{fig:ionization1} and Fig. \ref{fig:ionization2} that show that the mole fraction of electron is very small compared to the mole fraction of heavy particle in that range of temperature between $4000$ K to $6500$ K.


%

\subsection{Components of the generalized Ohm's law in a pore in the low Sun chromosphere}
Similarly as section~\ref{sec:pore}, we compute the components of the generalized Ohm's law from Eq.~\eqref{eq:ElectricField} using a Helium-Hydrogen mixture, from the simulation by Kitiashvili et al. \cite{Irina2}. According to the result found in Fig.~\ref{fig:ratiolambdae}, we assume an isotropic distribution of the transport properties.

\begin{figure}[!]
	\centering
	\includegraphics[trim={0cm 0 0cm 0},width=\columnwidth,clip]{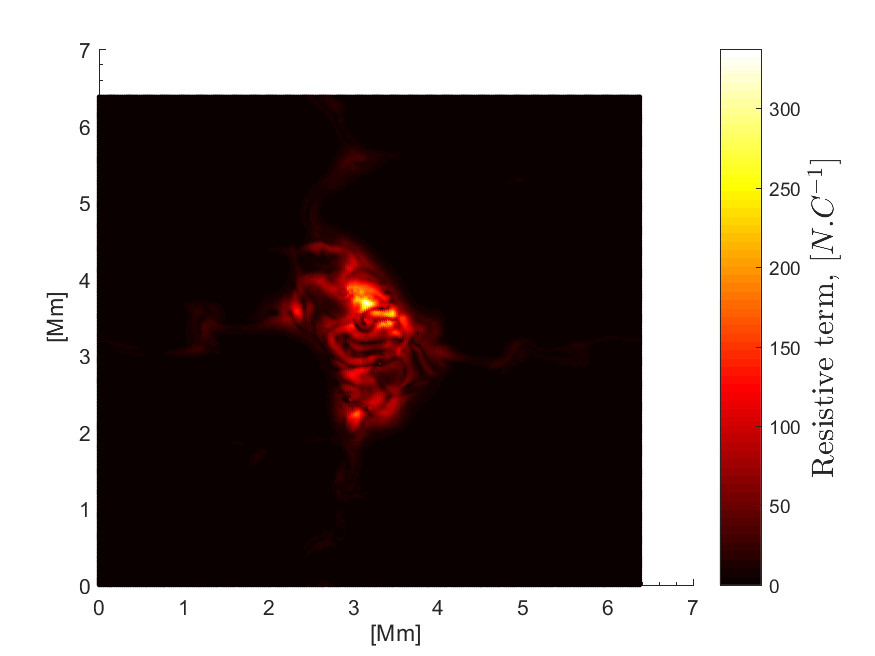}%
	\caption{Distribution of the resistive term (first term of (\ref{eq:ElectricField})), computed at the third order of the Laguerre-Sonine polynomials approximation, for the Helium-Hydrogen mixture $S_1$, based on the results of the radiative 3D MHD simulations of a pore by Kitiashvili et al.\cite{Irina2}} %
	\label{fig:resistiveterm}%
\end{figure}

\begin{figure}[!]
	\centering
	\includegraphics[trim={0cm 0 0cm 0},width=\columnwidth,clip]{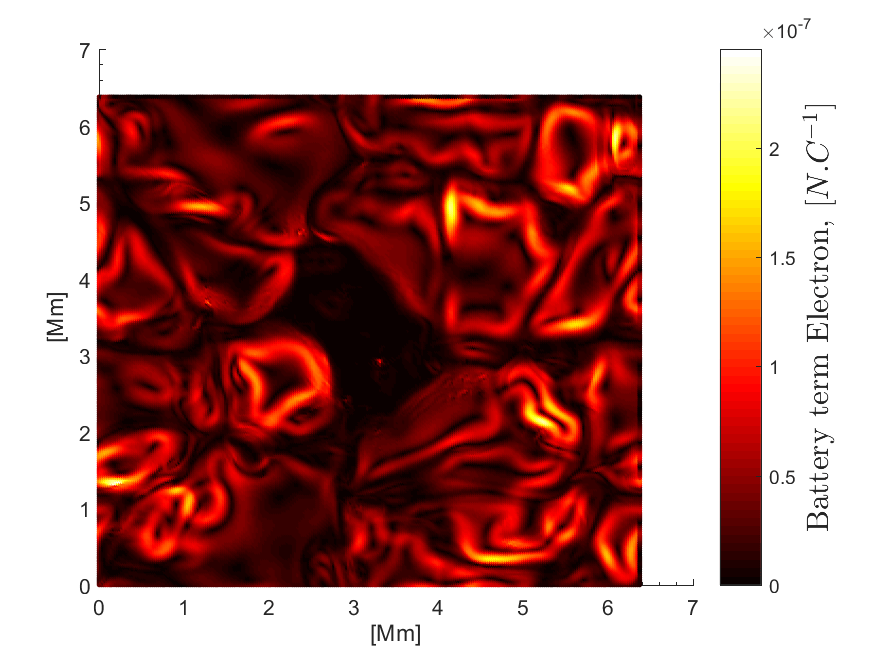}
	\caption{Distribution of the electron battery term (second term of (\ref{eq:ElectricField})), computed at the third order of the Laguerre-Sonine polynomials approximation, for the Helium-Hydrogen mixture $S_1$, based on the results of the radiative 3D MHD simulations of a pore by Kitiashvili et al.\cite{Irina2}} %
	\label{fig:batterytermelectron}%
\end{figure}

\begin{figure}[!]
	\centering
	\includegraphics[trim={0cm 0 0cm 0},width=\columnwidth,clip]{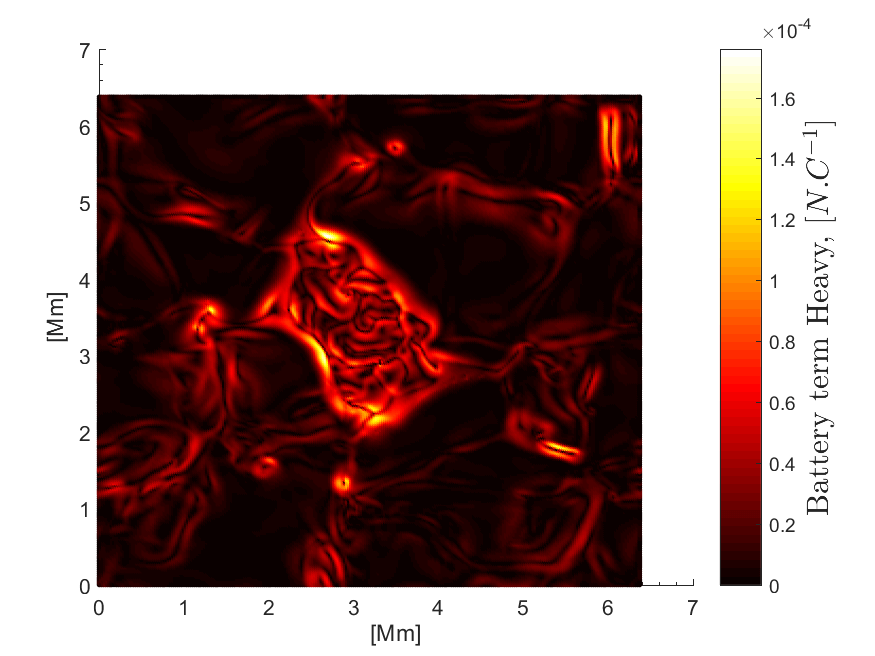}%
	\caption{Distribution of the heavy battery term (third term of (\ref{eq:ElectricField})), computed at the third order of the Laguerre-Sonine polynomials approximation, for the Helium-Hydrogen mixture $S_1$, based on the results of the radiative 3D MHD simulations of a pore by Kitiashvili et al.\cite{Irina2}} %
	\label{fig:batterytermheavy}%
\end{figure}

\begin{figure}[!]
	\centering
	\includegraphics[trim={0cm 0 0cm 0},width=\columnwidth,clip]{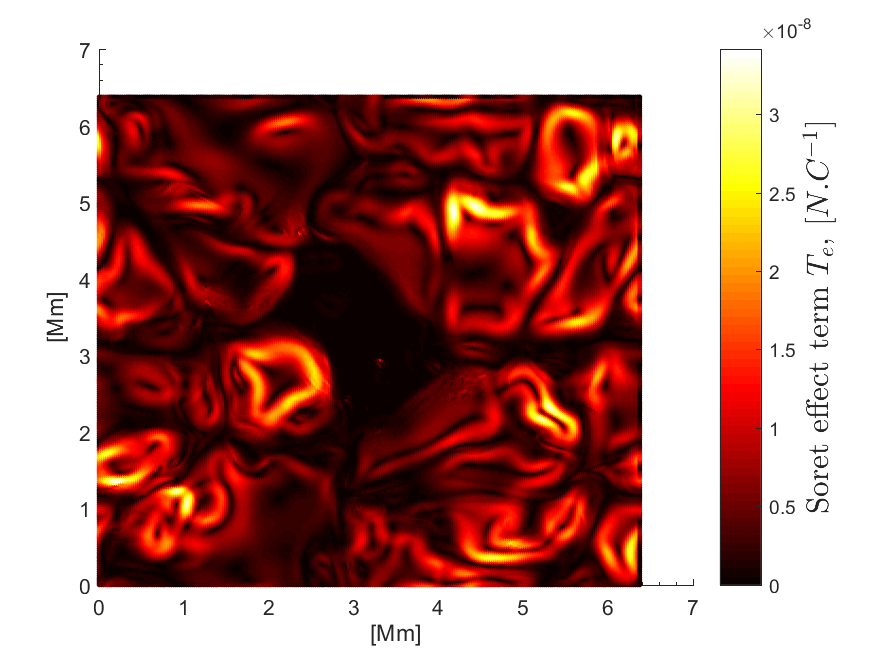}%
	\caption{Distribution of the Soret/Dufour term (fourth term of (\ref{eq:ElectricField})), computed at the third order of the Laguerre-Sonine polynomials approximation, for the Helium-Hydrogen mixture $S_1$, based on the results of the radiative 3D MHD simulations of a pore by Kitiashvili et al.\cite{Irina2}} %
	\label{fig:soreteffectte}%
\end{figure}

\begin{figure}[!]
	\centering
	\includegraphics[trim={0cm 0 0cm 0},width=\columnwidth,clip]{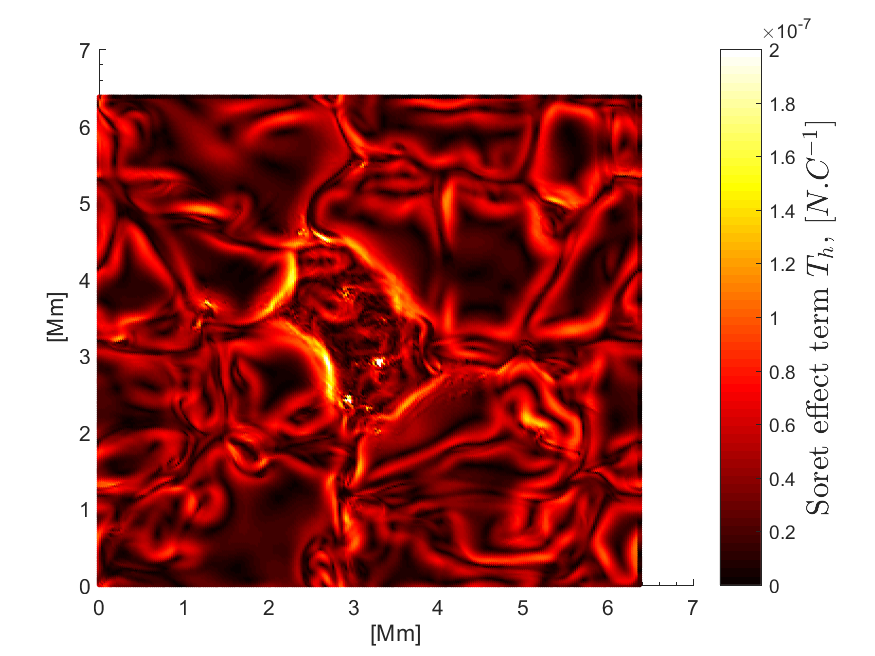}%
	\caption{Distribution of the Soret/Dufour term (last term of (\ref{eq:ElectricField})), computed at the third order of the Laguerre-Sonine polynomials approximation, for the Helium-Hydrogen mixture $S_1$, based on the results of the radiative 3D MHD simulations of a pore by Kitiashvili et al.\cite{Irina2}} %
	\label{fig:soreteffectth}%
\end{figure}

Fig.~\ref{fig:resistiveterm},\ref{fig:batterytermelectron},\ref{fig:batterytermheavy},\ref{fig:soreteffectte},and \ref{fig:soreteffectth} show the distribution of the resistive term, the electron battery term, the heavy particle battery term and the Soret terms for electron and heavy particle respectively. 

Under this condition, the dynamic of the electric field is dominated by the resistive term at the middle of the pore. The battery term for heavy particles appears to be higher at the middle of the pore and at some intergranulation boundaries. Finally, the Soret and battery terms for electrons have higher magnitude at the intergranulation boundaries, the latters is negligible compared to the other terms of the generalized Ohm's law. Indeed, this is due to the mole fraction of electrons which is very small compared to heavy particles under these conditions. These results are coherent with the distribution mole fraction presented in Fig.~\ref{fig:ionization1},Fig.~\ref{fig:ionization2} and Fig.~\ref{fig:ionization3}.

We have been able to compute all the components of the generalized Ohm's law applied to a pore. One can identify the distribution and the magnitude of the different terms. Under these conditions, since the ionization level is weak, the resistive term and battery term for heavy particles appear to dominate the dynamic of the electric field.


\section{Conclusion}

The present model, derived from the kinetic theory \cite{graille}, is neither a single-fluid MHD nor a multi-fluid model. It is an intermediary model between the single-fluid MHD models and the multi-fluid plasma models, i.e., a multi-component model. The main difference between the presented model \cite{graille} and the conventional multi-fluid approach that are derived from the kinetic theory is the scaling that has been used in the generalized Chapman-Enskog expansion. The scaling leads to a thermal non-equilibrium multicomponent model with one momentum equation, where the electrons diffuse in the heavy particle reference frame. These developments that are derived by Graille et al.~\cite{graille} lead to a model with an extended range of validity for partially and fully ionized plasma, non-,weakly- and strongly-magnetized plasmas and for a general multicomponent mixtures, which can be applied in Sun chromosphere conditions. From the set of governing equations, a generalized Ohm's law has been derived. A general expression of the resistive term as well as the battery term, has been obtained for a general multicomponent plasma. This general expression of the electric field can be simplified in a fully ionized plasma case (See Appendix B).

General conditions representative of the Sun chromosphere have been chosen in order to compute all the transport properties for a Helium-Hydrogen mixture $S_1$. A spectral Galerkin method based on a a third order of Laguerre-Sonine polynomials approximation has been implemented into Mutation++, an open-source library. First, in order to validate the model and the presented method, a comparison with the model of Braginskii \cite{Braginskii} has been performed in the case of a fully ionized plasma $S_2$. Both models are derived from the kinetic theory based a Chapman-Enskog expansion. Although differences are observed in 1-the structure of the governing equations and 2- the nature of the collision operators, the method that is used for computing the transport properties is the same in both cases. The corresponding integro-differential systems that allow for computing the transport properties are identical in both models. While in \cite{Braginskii}, the heavy transport properties are anisotropic, in the present model, the latter are isotropic. Nevertheless, under the studied chromospheric conditions both behave as isotropic. In \cite{Braginskii}, the corresponding series of Laguerre-Sonine polynomials are truncated at the second order \cite{balescu}, whereas a third order has been performed in the presented model. Good agreement has been obtained for the considered fully ionized $S_2$ mixture, in the chosen conditions.

Finally, by using Mutation++ Library, the presented method gives the possibility to compute all the transport properties for a partially ionized plasma for a given mixture. The obtained results strongly depend on the mole fraction and the ionization reactions between the species of the mixture. We have been able to identify the behavior of the transport coefficients related to the chemistry of the species in the partially ionized mixture $S_1$. We have computed the transport coefficient based on the results from a radiative 3D MHD simulations of a pore, in the highly turbulent upper layer of the solar convective zone by Kitiashvili et al.~\cite{Irina2}. In such conditions, the electrons are not magnetized, and the heavy particles dominate the dynamics of the pore. In the middle of the pore, the total heat flux is mainly due to the heavy particle flux, where the main components of the thermal conductivity are the heavy thermal conductivity and the reactive thermal conductivity. Similarly, we have been able to compute all the components of the generalized Ohm's law in the same conditions. The results show that the resistive term is dominating mainly the dynamic of the electric field at the pore. The battery term for heavy particle appears to be higher at the pore and some intergranulation boundaries. The rest of the terms appear to be negligible since the ionization level is small in such conditions.

\begin{acknowledgements}
	The research of Q. Wargnier is funded by an Idex Paris-Saclay interdisciplinary IDI PhD grant, and relies on the the support of NASA Ames Research Center (ARC), Advanced Supercomputing Division, von Karman Institute for Fluid Dynamics, CMAP - Initiative HPC@Maths from Ecole Polytechnique 
and  of Ecole doctorale de Math\'ematiques Hadamard. Part of this work was conducted during the 2018 NASA Summer Program at ARC. We would like to thank I.N. Kitiashvili for providing us with the data in order to evaluate the transport properties in subsection 5.2.
\end{acknowledgements}



\bibliographystyle{unsrt}
\bibliography{references}

\appendix

\section{Non-dimensional Boltzmann equations}
In Graille et al.~\cite{graille}, the non-dimensional Boltzmann equations for electron and heavy particles is
\begin{equation}
\begin{split}
\dt \fe&+\frac{1}{\epsilon \Mh}\left(\Ce+ + \epsilon \Mh \vitesse\right)\dscal \dx \fe \\&+\epsilon^{-(1-b)}\qe\left(\Ce+\epsilon \Mh \vitesse\right)\otimes \B \dscal \partial_{\Ce} \fe \\&+   \left(\frac{1}{\epsilon \Mh} \qe \E - \epsilon \Mh \frac{D \vitesse}{D t}\right)\dscal \partial_{\Ce} \fe\\&- (\partial_{\Ce} \fe\otimes\Ce)\pmat\dx \vitesse=\frac{1}{\epsilon^2}\mathcal{J}_{\elec},
\label{boltze}
\end{split}
\end{equation}
and for each heavy species, 
\begin{equation}
\begin{split}
\dt \fii&+\frac{1}{\Mh}\left(\Ci+\Mh \vitesse\right)\dscal\dx \fii\\&+\epsilon^{-(1+b)}\qi\left(\Ci+ \epsilon \Mh \vitesse\right)\otimes \B \dscal \partial_{\Ci} \fii \\&+ (\frac{1}{\epsilon \Mh} \frac{\qi}{\mi} \E -  \Mh \frac{D \vitesse}{D t}). \partial_{\Ci} \fii\\&-\left(\partial_{\Ci} \fii\otimes \Ci\right)\pmat\dx \vitesse=\frac{1}{\epsilon}\mathcal{J}_{{i}},\quad i \in \lourd.
\label{boltzi}
\end{split}
\end{equation}
Where the collision operators are defined as 
\begin{equation}
\mathcal{J}_\elec=\mathcal{J}_{\elec\elec}(\fe,\fe)+\sum_{j\in \heavy}\mathcal{J}_{\elec{j}}(\fe,\fj)., 
\label{collision1}
\end{equation}
\begin{equation}
\mathcal{J}_{i}=\frac{1}{\epsilon}\mathcal{J}_{{i}\elec}(\fii,\fe)+\sum_{j\in \heavy}\mathcal{J}_{{i}{j}}(\fii,\fj),\quad i\in\lourd.
\label{collision2}
\end{equation}
Where $\Ci$, $i \in \lourd$ and $\Ce$ are the peculiar velocities for heavy species and electron respectively. 

\section{Ohm's law for a fully ionized plasma $S_1$}
In a fully ionized plasma, the definition of the current density is 
\begin{equation}
\courantel=\ntot \qtot \vitesse+ \JJe 
\label{currentdensity}
\end{equation}
where the electron current density is defined by Eq.~ \eqref{eq:totalcurrent}. It should be noted that in the
particular case of a fully ionized plasma, no diffusion velocities of heavy species are considered, because the presented model is in the reference frame of heavy particles. \\

Then, similarly as the general case, the total current density can be expressed in function of fluxes, as follows, 
\begin{multline}
\courantel = \ntot \qtot \vitesse + \frac{\left(\nee\qe\right)^2 }{\pree}\mDee\Ep  - \nee\qe\Bigg[\mDee\frac{\dx \pree}{\pree} + \mDee\mchie\glogTe \Bigg] 
\label{eq:currentTermsNormalized2}
\end{multline}
It should be noted that the multicomponent electromagnetic matrices in a fully ionized plasma are 
\begin{align}
\matEp&= \mDee,
 \label{eq:electromagneti12} \\
\matpee  &= \mDee, \label{eq:electromagneti22}\\
\matpj&= 0, j\in\lourd\\
\matTe &= \mDee\mchie \label{eq:electromagneti32}\\
\matTh&=0. 
\label{eq:electromagneti42}
\end{align}

Using the Maxwell-Ampere's law $\dx \pvect \B = \mu_0 \courantel$ in a non-relativistic context, the expression of the electric field as a function of fluxes can be obtained
\begin{equation}
\E=-\vitesse \pvect\B +\bar{\bar\eta}_e \JJe  +\frac{\dx \pree}{\nee \qe}+\frac{\pree}{\nee \qe}\mchie\glogTe 
\label{ohmslaw}
\end{equation}
where the electron resistivity tensor is $\bar{\bar\eta}_\elec=\mDee^{-1}\pree/(n_\elec q_\elec)^2$.
\end{document}